\documentclass[%
 aip,
 cha,%
 amsmath,amssymb,
 reprint,%
]{revtex4-1}

\usepackage{hyperref}
\usepackage{graphicx}
\usepackage{dcolumn}
\usepackage{bm}
\usepackage{color,todonotes}
\usepackage[utf8]{inputenc}

\usepackage{xr}

\begin{document}


\title{{\small published as Kraemer et al., Chaos 28, 085720 (2018)\\ \vspace{12pt}}
Recurrence threshold selection for obtaining robust recurrence characteristics in different embedding dimensions}

\author{K.~Hauke Kr\"amer}
\email{hkraemer@pik-potsdam.de, hkraemer@uni-potsdam.de}
\affiliation{Potsdam Institute for Climate Impact Research, Telegrafenberg A31, 14473 Potsdam, 
Germany, EU
}%
\affiliation{%
Institute of Earth and Environmental Science, University of Potsdam, Karl-Liebknecht-Str. 
24-25, 14476 Potsdam-Golm, Germany, EU
}%

\author{Reik V. Donner}%
\affiliation{Potsdam Institute for Climate Impact Research, Telegrafenberg A31, 14473 Potsdam, 
Germany, EU
}%

\author{Jobst Heitzig}%
\affiliation{Potsdam Institute for Climate Impact Research, Telegrafenberg A31, 14473 Potsdam, 
Germany, EU
}%

\author{Norbert Marwan}
\affiliation{Potsdam Institute for Climate Impact Research, Telegrafenberg A31, 14473 Potsdam, 
Germany, EU
}%

\date{\today}

\begin{abstract}
The appropriate selection of recurrence thresholds is a key problem in applications of recurrence quantification analysis (RQA) and related methods across disciplines. Here, we discuss the distribution of pairwise distances between state vectors in the studied system's state space reconstructed by means of time-delay embedding as the key characteristic that should guide the corresponding choice for obtaining an adequate resolution of a recurrence plot. Specifically, we present an empirical description of the distance distribution, focusing on characteristic changes of its shape with increasing embedding dimension. 
Our results suggest that selecting the recurrence threshold according to a fixed percentile of this distribution reduces the dependence of recurrence characteristics on the embedding dimension in comparison with other commonly used threshold selection methods. Numerical investigations on some paradigmatic model system with time-dependent parameters support these empirical findings.
\end{abstract}


\maketitle

\begin{quotation}
Recurrence plots provide an intuitive tool for visualizing the (potentially multi-dimensional) trajectory of a dynamical system in state space. In case only univariate observations of the system's overall state are available, time-delay embedding has become a standard procedure for qualitatively reconstructing the dynamics in state space. The selection of a threshold distance $\varepsilon$, which distinguishes close from distant pairs of (reconstructed) state vectors, is known to have a substantial impact on the recurrence plot and its quantitative characteristics, but its corresponding interplay with the embedding dimension has not yet been explicitly addressed. Here, we point out that the results of RQA and related methods are qualitatively robust under changes of the (sufficiently high) embedding dimension only if the full distribution of pairwise distances between state vectors is considered for selecting $\varepsilon$, which is achieved by consideration of a fixed recurrence rate.      
\end{quotation}


\section{Introduction}\label{introduction}

A vector time series $\{\vec{x}_i\}_{i=1}^N$ (with $\vec{x}_i=\vec{x}(t_i)$) 
provides an approximation 
of a specific trajectory of a given dynamical system in finite-time and (for time-continuous 
dynamical systems) finite-resolution. In many real-world applications, however, inferring 
complete dynamical information from observations is hampered by the fact that only some of the 
dynamically relevant variables are directly observable. In 
such cases, it has been demonstrated\cite{Tak} that it is possible to qualitatively reconstruct 
representations of the unobserved components of a higher-dimensional system by means of 
embedding techniques applied to a suitably chosen individual component\cite{Letellier2002}. 
Specifically, time-delay embedding has become a widely utilized method in nonlinear time series 
analysis, where a series of univariate observations $\{x_i\}$ (the actual time series at hand) is unfolded into a sequence of 
$m$-dimensional state vectors $\{\vec{x}_i\}$\cite{Tak,Pac} defined as $\vec{x}_i=(x_i,x_{i-
\tau},\dots,x_{i-(m-1)\tau})^T$, where $m$ and $\tau$ denote the chosen embedding dimension and 
embedding delay, respectively.

Introduced by Eckmann et al.\cite{Eckmann}, recurrence plots (RPs) provide a versatile tool 
for visualizing and quantitatively analyzing the succession of dynamically similar states in a 
time series. For this purpose, dynamical similarity is measured in terms of some metric 
distance $d_{i,j}=\|\vec{x}_i-\vec{x}_j\|$ defined in the underlying system's (reconstructed) 
state space. Based on the resulting distance matrix $\mathbf{d}=(d_{i,j})$, a recurrence matrix 
$\mathbf{R}=(R_{i,j})$ is defined as a thresholded version such that its entries assume values of 1, if the distance between the two associated state vectors is smaller than or equal to a threshold $
\varepsilon$, and 0 otherwise:
\begin{equation}
R_{i,j}(\varepsilon)=\begin{cases}1:&d_{i,j} \leq \varepsilon \\
0:&d_{i,j} > \varepsilon,\end{cases} \qquad i,j = 1,...,N.
\end{equation}
Equivalently, we can write
\begin{equation}
R_{i,j}(\varepsilon)=\Theta (\varepsilon - d_{i,j}), \qquad i,j = 1,...,N,
\end{equation}
where $\Theta(\cdot)$ is the Heaviside function. In this definition, the threshold $\varepsilon$ is 
fixed with respect to all pairwise distances contained in $\mathbf{d}$.

\begin{figure*}
 \centering
 \includegraphics[width=0.875\textwidth]{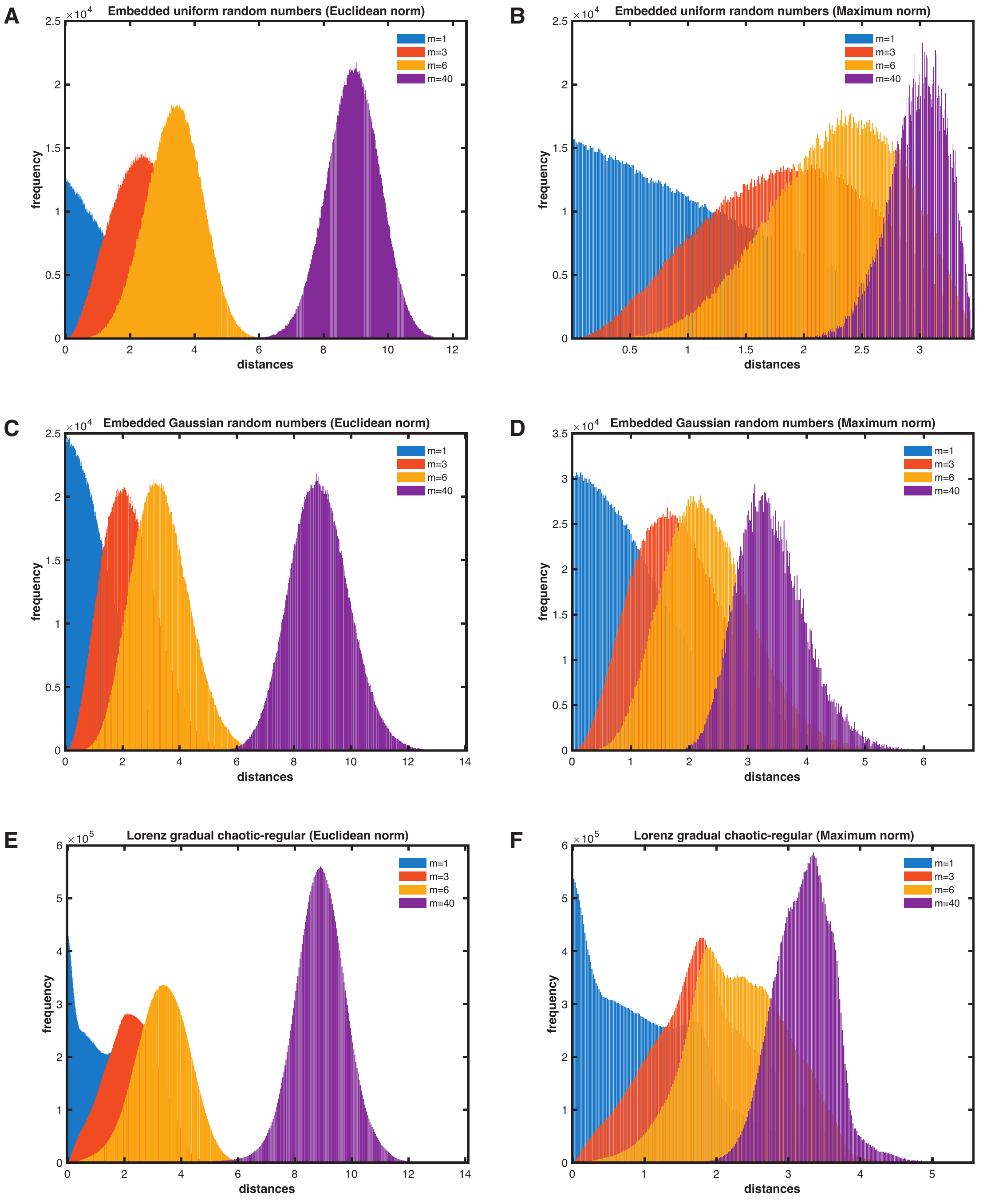}
\caption{Selected histograms of the $L_2$ (A,C,E) and $L_\infty$ (B,D,E) distances of $N=1,500$ independent random numbers with uniform (A,B) and Gaussian (C,D) distribution as well as (E,F) for the $y$ component of the Lorenz-63 system (Eq.~\eqref{Lorenz1}, $N=6,000$, see Section \ref{numerical}) with control parameters $\sigma=10$, $\beta=8/3$ and $r$ linearly increasing from 180 (chaotic regime) to 210 (periodic regime), for different embedding dimensions $m$.}
\label{dist_examples}
\end{figure*}

According to the above definition, for a given time series the recurrence matrix depends on the chosen recurrence threshold $\varepsilon$ together with the selected norm $\|
\cdot\|$ used for defining pairwise distances between the state vectors. In this work, we will 
restrict ourselves to two of the most commonly used norms: the Euclidean ($L_2$) and maximum ($L_\infty$, supremum, Chebychev) norms. Specifically, we will study how the distributions of 
pairwise $L_2$ and $L_\infty$ distances
depend on the embedding dimension.

Previous studies have provided various complementary suggestions for (i) selecting the right method of determining the recurrence threshold and (ii) choosing its actual value in some automatic way based on the specific properties of the system under study. Corresponding approaches include the spatial extent of the trajectory in the (reconstructed) state space\cite{Mind,Zbi1}, signal to noise 
ratio\cite{Zbi1,Zbi2,Thi1,schinkel2008}, the specific dynamical system underlying the time series under 
investigation\cite{Zbi2,Mat}, or properties of the associated recurrence 
network\cite{Marwan3,Don} with adjacency matrix $A_{i,j}=R_{i,j}-\delta_{i,j}$ (with $
\delta_{i,j}$ being the Kronecker symbol) like the percolation
threshold\cite{Donges2012,Jacob2016PRE}, 
second smallest eigenvalue of the graph's Laplacian\cite{Ero}, breakdown of $\varepsilon^{-1}$ 
scaling of the average path length\cite{Donges2012}, or information-theoretic 
characteristics\cite{Wie}. In practice, the appropriate choice of the method for determining the recurrence threshold, as well as its resulting value itself, can depend on the specific problem under study and take any of the above criteria or even some multiple-objective considerations based on different criteria into account. To this end, a general solution to the second problem of selecting a specific value of $\varepsilon$ has not yet been obtained, and we will also not address this problem specifically in the course of the present paper. Instead, we are attempting to provide some further insights into the first, more conceptual problem setting (i.e., which type of approach for selecting recurrence thresholds should be taken in case of varying situations such as different embedding dimensions).
\cite{Marwan3,Don,Donges2012,Wie}

As we will further detail in the course of this paper, some previously suggested approaches\cite{Koe,Mind,Zbi1,schinkel2008} to link a recurrence threshold to
a certain percentage of the maximal or mean distance of all pairwise distances 
of state vectors (i.e., a given fraction of the attractor's diameter in the reconstructed state space) cause the resulting recurrence characteristics to strongly depend on the embedding dimension. The reason for this behavior is that in addition to a 
general increase of distances\cite{Zimek} (depending on the chosen norm)\cite{Koe}, the shape of the distance distribution also changes with increasing embedding dimension (see Fig.~\ref{dist_examples} and further discussions in Section~\ref{influence}). 

It should be noted that embedding a time series with $\textit{m}\sim \mathcal{O}(10^1)$ or even larger can become necessary when the correlation dimension $D_2$ of the attractor is rather 
large. This is due to the fact that Takens' theorem (and several extensions thereof) guarantee 
the existence of a diffeomorphism between the original and the reconstructed attractor if 
\textit{m} satisfies $\textit{m}\geqslant 2 D_2 +1$\cite{Marwan1,Tak,Sau}. Hegger et al.
\cite{Heg} emphasize that it is also advisable to choose a rather high value of $m$ when 
dealing with time series originating from a $D$-dimensional deterministic system that is 
driven by $P$ slowly time dependent parameters. An appropriate yet conservative choice for $m$ then fulfills $m \geqslant 2(D+P)$. Concerning practical applications of nonlinear time series analysis, one commonly deals with signals originating from complex, non-stationary systems and, therefore, high embedding dimensions can become necessary, requiring threshold selection methods which lead to robust results of RQA and related state space based techniques that are robust under different choices of the embedding dimension.

In the following Section~\ref{influence}, we study the influence of an increasing embedding 
dimension on the shape of the distance distribution in more detail. We deduce that, in order to 
avoid problems arising due to an unfavorable fixed recurrence threshold when varying $m$, 
we could choose $\varepsilon$ as a certain percentile of the distance distribution rather than a certain percentage of the maximum or mean phase space diameter. Successively, Section~\ref{numerical} presents a numerical example of a classical Lorenz-63 system with a time-dependent parameter, illustrating that the changes in some recurrence characteristics with varying embedding dimension are particularly small under a fixed recurrence rate in comparison with other strategies. Finally, the main results of this study are summarized in Section~\ref{conclusion}.


\section{Influence of embedding dimension on the distance distribution}\label{influence}

Let us consider a univariate time series $\{x_i\}$ of length $N$. As an overarching question, we study the effect of time-delay embedding on the distribution between all pairwise distances of its reconstructed state vectors. The variation of this distribution with increasing embedding dimension $m$ is expected to depend on the chosen norm used for the calculation of distances. Note that the effective number of state vectors $N_{eff}(m)=N-(m-1)\tau$ available for estimating the probability distribution of distances in $m$ dimensions will decrease with $m$. In order to avoid sample size effects in comparing the results for different $m$, we therefore choose $N$ sufficiently large so that $1-N_{eff}(m_f)/N\ll 1$, where $m_f$ is the largest considered embedding dimension.


\subsection{Maximum norm}\label{influence_maximum}   

Numerical results for different types of systems demonstrate (see Appendix \ref{S_influence}) that the largest of all pairwise $L_\infty$ distances, $d_{max}^{(\infty)}$, stays constant with increasing embedding dimension, whereas the mean of all pairwise $L_\infty$ distances, $d_{mean}^{(\infty)}$, monotonically increases with $m$ (Fig.~\ref{S_figure1}). In order to 
understand this observation, recall that the $L_\infty$ distance between two embedded state 
vectors $\vec{x}_i = (x_{i,1}, x_{i,2},\ldots,x_{i,m})^T$ and $\vec{x}_j = (x_{j,1}, x_{j,2},\ldots,x_{j,m})^T$ is  
\begin{equation}
\|\vec{x}_i-\vec{x}_j\|_\infty = \max_{k=1,\dots,m} \left| x_{i,k}-x_{j,k} \right| = 
d^{(\infty)}_{i,j}(m)
\end{equation}
For $m=1$ (i.e., no embedding), the distance between two observations at times $t_i$ and $t_j$ 
therefore is simply $d^{(\infty)}_{i,j}(1)=\left|x_i-x_j\right|$. For $m=2$, we find
\begin{align}
d^{(\infty)}_{i,j}(2) &= \max \{\left|x_i - x_j\right|,\left|x_{i+\tau}-x_{j+\tau}\right|\} 
\notag \\
&= \max\left\{d^{(\infty)}_{i,j}(1),\left|x_{i+\tau}-x_{j+\tau}\right|\right\} \geqslant 
d^{(\infty)}_{i,j}(1).
\end{align}
By induction, we can easily show that
\begin{equation*}
d^{(\infty)}_{i,j}(m) = \max \left\{ d^{(\infty)}_{i,j}(m-1), \left|x_{i-(m-1)\tau}-x_{j-
(m-1)\tau}\right| \right\} 
\end{equation*}
and therefore
\begin{equation}
d^{(\infty)}_{i,j}(m) \geqslant d^{(\infty)}_{i,j}(m-1) \qquad \forall\ m>1.
\end{equation}
Hence, considering all possible pairs of state vectors $(\vec{x}_i,\vec{x}_j)$ from the time
series, the largest $L_\infty$ distance 
\begin{equation}
d^{(\infty)}_{max}(1)= \max_{i,j}[d^{(\infty)}_{i,j}(1)] = \max_{i,j}[d^{(\infty)}_{i,j}(m)] = d_{max}^{(\infty)}(m) \qquad \forall\ m \notag
\end{equation}
\noindent
cannot change with $m$, since the largest maximum distance will already appear for $m=1$. The mean distance 
\begin{equation}
d_{mean}^{(\infty)}(m) = \frac{1}{N_{{eff}}^2(m)} \sum_{i,j=1}^{N_{{eff}}(m)} d_{i,j}^{(\infty)}(m), 	\notag
\end{equation}
\noindent
however, necessarily increases with $m$ or stays at most constant. More specifically, as $m$ increases, smaller distances systematically disappear, so that 
the entire distribution is systematically shifted towards its (constant) maximum, thereby 
becoming narrower and exhibiting an increasing mean along with decreasing variance. We conjecture 
that, for large $m$, the distribution of $d^{(\infty)}(m)$ will converge to a limiting 
distribution (see below) possibly depending on the embedding delay $\tau$.


\subsection{Euclidean norm}\label{influence_euclidean}

In case of the \textit{$L_2$} (Euclidean) norm, both mean and maximum of all pairwise distances ($d^{(2)}_{mean}(m)$ and $d^{(2)}_{max}(m)$, respectively) monotonically 
increase with rising $m$ (Appendix \ref{S_influence}, Fig.~\ref{S_figure2}). This can be understood as follows: The $L_2$ 
distance between two points in an $m$-dimensional state space, $\vec{x}_i$ and $\vec{x}_j$, is 
given as  
\begin{equation}
\|\vec{x}_i-\vec{x}_j\|_2 = \biggl( \sum_{k=1}^{m}\left|x_{i,k}-x_{j,k}\right|
^2\biggr)^{\frac{1}{2}} = d_{i,j}^{(2)}(m) \label{euc1}
\end{equation}
For the squared $L_2$ distance, this implies:
\begin{align}
\left[d^{(2)}_{i,j}(1)\right]^2 & = (x_i -x_j)^2 \notag \\
\left[d^{(2)}_{i,j}(2)\right]^2 & = (x_i -x_j)^2+(x_{i-\tau} -x_{j-\tau})^2 \notag \\
\label{euc2}&= \left[d^{(2)}_{i,j}(1)\right]^2 +(x_{i-\tau} -x_{j-\tau})^2 \notag \\
&\geqslant \left[d^{(2)}_{i,j}(1)\right]^2 \\
&\vdots \notag\\
\label{euc3} \left[d^{(2)}_{i,j}(m+1)\right]^2&\geqslant \left[d^{(2)}_{i,j}(m)\right]^2 
\geqslant \dots \geqslant \left[d^{(2)}_{i,j}(1)\right]^2,
\end{align}
which explains the observed behavior of both mean and maximum distance using the $L_2$ norm. Specifically, unlike 
for $L_\infty$, the maximum $L_2$ distance between two points is not bound by the largest 
pairwise distance in one dimension. 

In a similar way, we may argue for all $L_p$ distances ($p\in(0,\infty)$) defined as
\begin{equation}
\|\vec{x}_i-\vec{x}_j\|_p = \biggl( \sum_{k=1}^{m}\left|x_{i,k}-x_{j,k}\right|
^p\biggr)^{\frac{1}{p}} = d_{i,j}^{(p)}(m)
\label{Lp_distances}
\end{equation}
that, by the same argument as above,
\begin{equation}
\left[d^{(p)}_{i,j}(m+1)\right]^p \geqslant \left[d^{(p)}_{i,j}(m)\right]^p,
\end{equation}
implying again a monotonic increase of mean and maximum distances with rising embedding dimension (recall the positive semi-definiteness of distances and $p$).


\subsection{Changing shape of distance distribution with increasing embedding dimension}
\label{influence_shape}

Building upon our previous considerations and numerical results, a mathematically more specific yet challenging question is how exactly an increasing 
embedding dimension $m$ is affecting the shape of the distribution of all pairwise distances 
rather than just its central tendency (mean). 

For the maximum norm, one may argue that the individual components of each embedded state 
vector are commonly constructed such that they are as independent as possible\cite{Fraser}. Accordingly, for 
a system without serial correlations (i.e., uncorrelated noise), the absolute differences 
$d=d^{(\infty)}(1)$ between the components of two state vectors are also independent, 
identically distributed (i.i.d.) and lie within the interval $[0,d_{max}]$.
In 
such case, for sufficiently large $m$, the pairwise $L_\infty$ distance between two state 
vectors can be interpreted as the 
maximum of $m$ i.i.d.~variables that are bounded from above, which should 
follow a reversed Weibull distribution according to the Fisher-Tippett-Gnedenko theorem from extreme value statistics. Note, however, that this expectation is valid only if $m$ is sufficiently large and the i.i.d.~assumption is (approximately) fulfilled, both of which do not 
necessarily have to be the case for real-world time series. Moreover, it is not guaranteed that the given distance distribution in one dimension lies within the domain of attraction of the reversed Weibull class\cite{Leadbetter}, which calls for further theoretical investigation in each specific case.

For other $L_p$ norms including the Euclidean norm, the aforementioned considerations do not apply. Instead, for any $L_p$ norm with $p<\infty$,

\begin{itemize}
 \item the pairwise distances $d^p$ are of the form $(\sum_i z_i^p)^{1/p}$ ($i=1,...,m$) as given in Eq.~\eqref{Lp_distances} with approximately i.i.d.\ variables $z_i$.
 
 \item From the central limit theorem it follows that the distribution of $d^p$ is approximately a normal distribution with mean and standard deviation growing proportionally with $m$ and $\sqrt{m}$, respectively, for large $m$.
 
 \item The coefficient of variation of $d^p$ thus declines approximately as $\sim 1/\sqrt{m}$.

 \item For large $m$, also $d=(d^p)^{1/p}$ is approximately normally distributed with mean and standard deviation growing approximately as $\sim m^{1/p}$ and $\sim\sqrt{m}\frac{dz^{1/p}}{dz}|_{z=m}\sim\sqrt{m}m^{1/p - 1}=m^{1/p - 1/2}$. 
 
 \item The coefficient of variation of $d$ thus behaves approximately as $\sim m^{1/p - 1/2}/m^{1/p} = 1/\sqrt{m}$, just as for $d^p$.
 
 \item As a consequence, the relative variability of $d$ narrows in the same fashion for all $p<\infty$ as $m$ grows, and only the growth of the absolute scale of $d$ with $m$ depends on $p$ (``curse of dimensionality''\cite{Zimek}). 
\end{itemize}

The considerations made above do explain the numerical results in
Fig.~\ref{dist_examples}, showing histograms of the distances of three different time series for selected values of the embedding dimension $m$ and for the $L_2$ and $L_{\infty}$ norms. In addition to time series fulfilling the i.i.d.~assumption (Fig.~\ref{dist_examples} A,B,C,D), here we are also interested in deterministic systems. As an illustrative example, we choose the Lorenz-63 system (Eq.~\eqref{Lorenz1}, Fig.~\ref{dist_examples} E,F) in some non-stationary (drifting parameter) setting, which will be further studied in Section \ref{numerical}.

In this regard, it is confirmed that the expectation value of the distance distribution takes higher values with increasing $m$. The probability to find small distances therefore decreases. In case of the $L_{\infty}$ norm (Fig.~\ref{dist_examples} B,D,F), this growth is bounded and we can identify a convergence of the distribution, in some cases eventually towards the aforementioned reversed Weibull distribution. In turn, for the $L_2$ norm (Fig.~\ref{dist_examples} A,C,E) the convergence towards a normal distribution is discernible. Considering the Lorenz-63 time series (Fig.~\ref{dist_examples} E,F), the empirical expectations are approximately met by the observations, even though the distribution of $L_\infty$ distances exhibits a slightly more complex (i.e., less symmetric) shape than for the two noise series. Specifically, for the $L_2$ norm the resulting distance distribution is left-skewed with a pronounced lower tail (see Fig.~\ref{dist_examples} E), whereas for the $L_\infty$ norm we rather observe a disturbed Weibull-like shape. Notably, the i.i.d.~assumption is violated when dealing with such a deterministic dynamical system. For a more detailed characterization of the shape of the empirically observed pairwise distance distributions shown in Fig.~\ref{dist_examples}, see Appendix \ref{S_variations}.

In general, we emphasize that it is not straightforward to analytically describe the shape of the distance distribution of an embedded time series stemming from an arbitrary dynamical system with potentially nontrivial serial correlations. Regarding our overarching question how we could automatically choose a recurrence threshold such that the resulting recurrence characteristics are as independent as possible of the embedding dimension and chosen norm, we need to consider both,  
\begin{itemize}
\item[(i)] the general increase of distances together with their successive concentration and
\item[(ii)] the varying shape of the distribution of distances 
\end{itemize}
with increasing embedding dimension. The first aspect could be accounted for by relating the threshold selection to the spatial extent of the state space object (attractor), similar as, for instance, suggested by Abarbanel\cite{Abarbanel} in the context of the false nearest neighbor algorithm. However, our findings suggest that accounting for the second point is key to an appropriate recurrence threshold selection method that relieves the effects of the embedding dimension on the recurrence properties as much as possible. As a simple possible solution, we recommend to use a numerical estimate of a certain (sufficiently low) percentile of the distance distribution as threshold\cite{Marwan3,Don,Donges2012,Wie}. This approach considers both above mentioned effects and leads to a constant global recurrence rate (which equals the chosen percentile). As a result, the recurrence properties become much less dependent on the embedding dimension and chosen norm than when using other methods, as we will exemplify in the following section.    

We emphasize that in addition, by conserving the recurrence rate, possible dependences of RQA characteristics on the density of recurrences for different $m$ are omitted, and corresponding residual changes of these measures upon varying $m$ could rather point to either insufficiently low embedding dimension (missing essential factors contributing to the system's dynamics, in a similar spirit as, e.g., for the false nearest neighbor method) or spurious recurrence structures arising from overembedding\cite{Thiel2006}. These ideas should be further studied in future work.


\section{Numerical example}\label{numerical}

In this section, we will demonstrate the effect of the varying shape of the distance distribution with increasing embedding dimension on different threshold selection approaches working with a globally fixed value of $\varepsilon$. 
In order to mimic a practically relevant test case of a non-stationary low-dimensional dynamical system, where we should use some higher embedding dimension (following \citeauthor{Heg}\citep{Heg}) instead of a more moderate choice, we consider the classical Lorenz-63 system\cite{Lorenz}
\begin{equation}
\begin{array}{rcl}
\dot{x}&=&\sigma(y-x) \\
\dot{y}&=&x(r-z)-y \\
\dot{z}&=&xy - \beta z.
\end{array}
\label{Lorenz1}
\end{equation}

Depending on the parameters $\sigma$, $\beta$ and $r$, the system exhibits either regular or chaotic dynamics. Here, we consider a transitory setting, where the parameter $r$ gradually increases from 180 to 210 while keeping $\beta=8/3$ and $\sigma=10$ fixed. In this case, the system undergoes a transition from a chaotic regime into a regular (limit cycle) phase as $r$ rises before it exhibits again a chaotic behavior. Note again that instead of studying the stationary Lorenz-63 system for different values of $r$, we intentionally employ a gradual parameter change leading to a non-stationary system which calls for a systematic overembedding when performing nonlinear time series analysis\citep{Heg}. Specifically, we implement a linear variation of $r$ as
\begin{equation}
r(t_{is}) = 180 + 2.5\cdot 10^{-2}t_\text{is}.
\label{parameter_change}
\end{equation} 

For numerically solving this system of equations, we use a fourth-order Runge-Kutta integrator 
with an integration step of $t_{is} 
= 0.001$ and a total of 1,300,000 iterations. Therefore, we simulate the system's evolution over 1,300 time units (t.u.). By using a sampling interval of $\delta t = 0.2$ t.u.~we obtain 6,500 samples forming our time series for the three components $x$, $y$ and $z$.
We remove the first 500 points ($\widehat{=}100$~t.u.) that could be affected by transient dynamics and 
retain the remaining 6,000 points ($\widehat{=}1200$~t.u.) of the $y$ component for further analysis.

We integrate the Lorenz-63 equations, Eq.~\eqref{Lorenz1}, with the linear parameter change, Eq.~\eqref{parameter_change}, 1,000 times with randomly chosen initial conditions, embed the $y$ component time series using a delay $\tau=4$, consistent with the first local minimum of the mutual information\cite{Fraser}, and assess the resulting RPs. For each of these 1,000 RPs, we use a running window along their main diagonal with a window size of $w=400$ and mutual shift of $ws=40$ data points, i.e., 90\% overlap between consecutive windows, to quantitatively study the time-dependence of the resulting recurrence characteristics. We repeat this procedure for embedding dimensions ranging from $m=3$ to $m=10$ and for four different threshold selection methods: (i) a fixed percentile of the distance distribution (as recommended by our theoretical considerations in Section~\ref{influence}) as well as some fixed percentage of the (ii) maximum, (iii) mean and (iv) median pairwise distances between all state vectors in the reconstructed state space, respectively. 

Since we are aiming to study the change of recurrence properties associated with a transition between chaotic and periodic dynamics and vice versa, we choose the \textit{recurrence time entropy (RTE)}. Here, instead of using the diagonal or vertical ``black'' (recurrence) lines in the RP as in most ``conventional'' RQA measures, we use ``white'' (non-recurrence) vertical lines with lengths $t_w$, as they correspond to recurrence times. In general, such recurrence times can be estimated directly from the RP in different ways \cite{ngamga2016}, among which the vertical non-recurrence lines offer a particularly simple estimator. The normalized entropy of the distribution of recurrence times, referred to as the \textit{recurrence period density 
entropy}\cite{little2007} and originally introduced without any direct link to RPs, is given as
\begin{equation}\label{eq_rte}
RTE = -\frac{1}{\ln T_\text{max}}\sum_{t_w=1}^{T_\text{max}} p(t_w) \ln p(t_w) \in [0,1] 
\end{equation}
with $p(t_w)$ being the probability of a recurrence time $t_w$ and $T_{max}$ the largest recurrence time. Using RPs, it is possible to estimate $p(t_w)$ from the histogram of recurrence times, $h(t_w)$, as $p(t_w) = \frac{h(t_w)}{\sum_{t_w} h(t_w)}$, i.e., as the probability to find a white vertical line of exactly length $t_w$ in the RP. It can be shown that $RTE$ is closely linked to the Kolmogorov-Sinai (KS) entropy of the system under study~\cite{baptista2010}.

\begin{figure*}
 \centering
 \includegraphics[width=\textwidth]{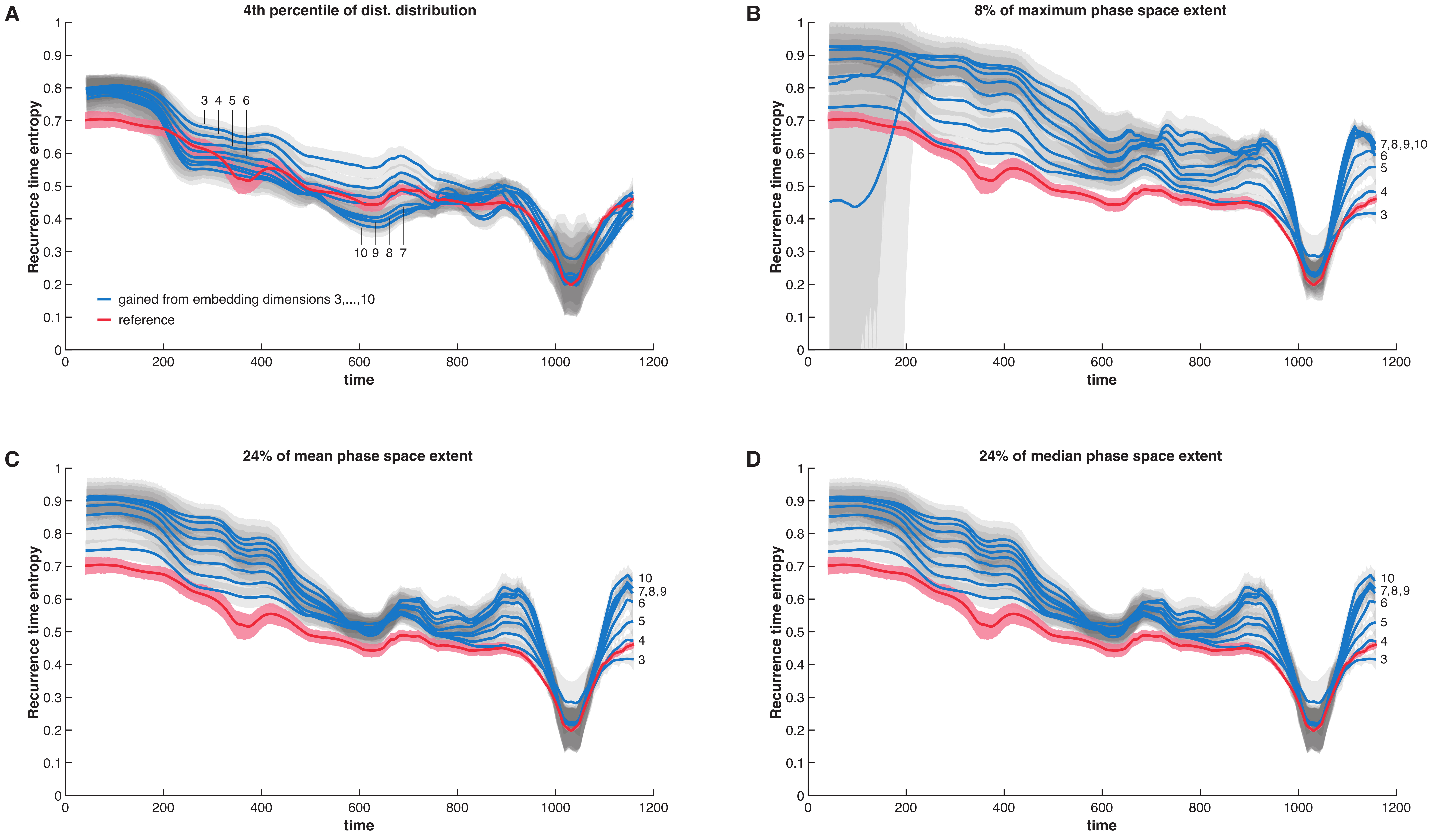}
\caption{Time-dependence of $RTE$ (ensemble means and two-sided 90\% confidence intervals from 1,000 independent realizations) based on the $y$ component of the non-stationary Lorenz-63 system (see text for details) using the $L_2$ norm. The blue lines show the results for time-delay embedding with different embedding dimensions ($m=3,\dots,10$) and for four different methods to select the recurrence threshold according to (A) a certain percentile of the distance distribution and some percentage of the (B) maximum, (C) mean and (D) median distance between state vectors on the reconstructed attractor. The actual threshold values ($4$th percentile, $8\%$, $24\%$ and $24\%$, respectively) have been chosen such that the global recurrence rate of approximately $4\%$ is achieved for each method in the embedding scenario with $m=3$. The red line shows the reference time series gained from 1,000 independent realizations of the non-stationary Lorenz-63 system by randomly choosing initial conditions and using all three components as state variables. Shaded areas (gray and red) indicate the two-sided $90\%$ confidence intervals estimated from the respective ensembles.}
\label{trans_statistic}
\end{figure*}

We choose the actual recurrence threshold for each threshold selection method (i)--(iv) such that a global recurrence rate of $RR\approx 4\%$ is achieved in all four cases for $m=3$. Therefore, for each embedding dimension we obtain a distribution of $1,000$ $RTE$ time series and show the mean (blue lines in Fig.~\ref{trans_statistic}) together with the two-sided 90\% confidence interval ($[5\%,95\%]$) (gray shaded areas). In order to put these time dependent $RTE$ estimates of the non-stationary Lorenz-63 system into a context, we consider a reference reflecting the time-dependent $RTE$ values directly computed from the true three-dimensional state vectors without embedding, using otherwise the same analysis strategy (window size and overlap) as for the embedding scenario. 
Thus, for each point in time we obtain 1,000 reference measurements and consider the mean (red line) and the two-sided 90\% confidence interval (red shaded area in Fig.~\ref{trans_statistic}).

The robustness of the observed time-dependence of $RTE$ with respect to the chosen embedding dimension when using a fixed percentile of the distance distribution (i.e., a fixed recurrence rate) is shown in Fig.~\ref{trans_statistic}A (here we used the $L_2$ norm, but the results are similar when using the $L_\infty$ norm). For any embedding dimension larger than $m=4$, the variations of the $RTE$ estimates originating from the embedding procedures match the red reference time series within its uncertainties for times $t\gtrsim 200$. For adequately revealing the chaotic regime in the first part until $t\approx 160$, an embedding dimension larger than $m=7$ seems to be inevitable, whereas results from any embedding dimension coincide with the reference estimate within its uncertainties at the limit cycle regime ($1,000\leq t \leq 1,080$). In case of not using the recommended threshold selection method, this robustness is clearly lost (Fig.~\ref{trans_statistic}B,C,D), and only the limit cycle regime (plus some shorter sections before) are properly revealed by the estimates obtained in the reconstructed state space.

Considering the results of Section \ref{influence}, the reason for the failure of the methods based on individual location parameters (maximum, mean, median) of the pairwise distance distribution between state vectors for higher embedding dimensions is the change in the shape of that distribution beyond its characteristic location and range parameters. Appendix \ref{S_Lorenz_exp} demonstrates this effect on the RPs in some more detail. Hence, we argue that selecting the recurrence threshold at some percentile of the distance distribution is to be preferred if we aim to obtain stable results for a broad range of embedding dimensions, which is the case if we wish to automatically choose fixed recurrence thresholds for the analysis of arbitrary complex systems. 

We note that the presented example has focused on a recurrence characteristic that is particularly well suited for detecting transitions between chaotic and periodic dynamics
and is linked to a dynamical invariant. Other recurrence characteristics, like classical RQA measures or recurrence network characteristics, have been found to exhibit less stable variations with changing embedding dimension (not shown) and are therefore not further discussed here. Clarifying the reasons for the different behaviors of different recurrence characteristics will be an important subject of future work.


\section{Conclusions}\label{conclusion}

We have discussed the changing shape of the distribution of pairwise distances between state vectors obtained by time delay embedding with increasing embedding dimension and its implications for different methods for selecting temporally fixed recurrence thresholds. While specific values of the recurrence threshold should always be chosen based upon a multitude of criteria ranging from time series length to different topological and/or geometric characteristics of the associated trajectory, we have provided both empirical arguments and numerical indications that selecting the recurrence threshold at a prescribed percentile of the distance distribution (i.e., conserving the global recurrence rate) results in quantitative recurrence characteristics that are more stable under changes of the embedding dimension than when using alternative approaches. 
In the latter context, we have demonstrated that measures from RQA and related frameworks may exhibit a crucial dependence on the embedding dimension when selecting the recurrence threshold according to a certain percentage of the mean or maximum state space diameter, as sometimes suggested in other works\cite{Zbi1,schinkel2008}. This also indicates that some alternative approaches, such as normalizing the time series and applying a uniform threshold independent of the embedding dimension and the considered norm\cite{Jacob2016PRE}, are not likely to perform well for any kind of data, when neglecting the effect on the distance distribution with increasing embedding dimension.

At the conceptual level, 
our general discussion of the changing shape of distance distributions with embedding dimension has led to some interesting follow-up questions associated with the convergence properties of these distributions at high embedding dimensions, which should be further addressed in future studies. Notably, the relationship between the distribution of $L_\infty$ distances and extreme value statistics clearly deserves further investigations to fully understand the emerging shape of the distributions as the embedding dimension becomes large. As a cautionary note, we emphasize that the considerations presented in this work relate exclusively to the concept of time delay embedding as the most widely applied embedding technique, but not necessarily to methodological alternatives like derivative embedding\cite{Lekscha}, for which the metric properties of different components of the embedding vector cannot be easily related to each other. 

Taken together, the results presented in this work are important for automatizing the problem of data-adaptive recurrence threshold selection, which is key for further widening the scope of applications of recurrence plots, recurrence quantification analysis and related techniques across scientific disciplines. Especially in the context of long time series originating from non-stationary systems, which frequently appear in many fields of science, a generally applicable approach is crucial for obtaining reliable and easily interpretable results.

\section*{Acknowledgments}
This work has been financially supported by the German Research Foundation (DFG projects no.~MA4759/8 and MA4759/9), the European Union's Horizon 2020 Research and Innovation Programme under the Marie Sk{\l}odowska-Curie grant agreement no.~691037 (project QUEST), and the German Federal Ministry of Education and Research via the Young Investigators Group CoSy-CC$^2$ (Complex Systems Approaches to Understanding Causes and Consequences of Past, Present and Future Climate Change, grant no.~01LN1306A).

\bibliography{refs_paper}

\begin{thebibliography}{31}%
\makeatletter
\providecommand \@ifxundefined [1]{%
 \@ifx{#1\undefined}
}%
\providecommand \@ifnum [1]{%
 \ifnum #1\expandafter \@firstoftwo
 \else \expandafter \@secondoftwo
 \fi
}%
\providecommand \@ifx [1]{%
 \ifx #1\expandafter \@firstoftwo
 \else \expandafter \@secondoftwo
 \fi
}%
\providecommand \natexlab [1]{#1}%
\providecommand \enquote  [1]{``#1''}%
\providecommand \bibnamefont  [1]{#1}%
\providecommand \bibfnamefont [1]{#1}%
\providecommand \citenamefont [1]{#1}%
\providecommand \href@noop [0]{\@secondoftwo}%
\providecommand \href [0]{\begingroup \@sanitize@url \@href}%
\providecommand \@href[1]{\@@startlink{#1}\@@href}%
\providecommand \@@href[1]{\endgroup#1\@@endlink}%
\providecommand \@sanitize@url [0]{\catcode `\\12\catcode `\$12\catcode
  `\&12\catcode `\#12\catcode `\^12\catcode `\_12\catcode `\%12\relax}%
\providecommand \@@startlink[1]{}%
\providecommand \@@endlink[0]{}%
\providecommand \url  [0]{\begingroup\@sanitize@url \@url }%
\providecommand \@url [1]{\endgroup\@href {#1}{\urlprefix }}%
\providecommand \urlprefix  [0]{URL }%
\providecommand \Eprint [0]{\href }%
\providecommand \doibase [0]{http://dx.doi.org/}%
\providecommand \selectlanguage [0]{\@gobble}%
\providecommand \bibinfo  [0]{\@secondoftwo}%
\providecommand \bibfield  [0]{\@secondoftwo}%
\providecommand \translation [1]{[#1]}%
\providecommand \BibitemOpen [0]{}%
\providecommand \bibitemStop [0]{}%
\providecommand \bibitemNoStop [0]{.\EOS\space}%
\providecommand \EOS [0]{\spacefactor3000\relax}%
\providecommand \BibitemShut  [1]{\csname bibitem#1\endcsname}%
\let\auto@bib@innerbib\@empty
\bibitem [{\citenamefont {Takens}(1981)}]{Tak}%
  \BibitemOpen
  \bibfield  {author} {\bibinfo {author} {\bibfnamefont {F.}~\bibnamefont
  {Takens}},\ }\bibfield  {title} {\enquote {\bibinfo {title} {Detecting
  strange attractors in turbulence},}\ }\href@noop {} {\bibfield  {journal}
  {\bibinfo  {journal} {Lecture Notes in Mathematics}\ }\textbf {\bibinfo
  {volume} {898}},\ \bibinfo {pages} {366–381} (\bibinfo {year}
  {1981})}\BibitemShut {NoStop}%
\bibitem [{\citenamefont {Letellier}\ and\ \citenamefont
  {Aguirre}(2002)}]{Letellier2002}%
  \BibitemOpen
  \bibfield  {author} {\bibinfo {author} {\bibfnamefont {C.}~\bibnamefont
  {Letellier}}\ and\ \bibinfo {author} {\bibfnamefont {L.~A.}\ \bibnamefont
  {Aguirre}},\ }\bibfield  {title} {\enquote {\bibinfo {title} {Investigating
  nonlinear dynamics from time series: The influence of symmetries and the
  choice of observables},}\ }\href {\doibase 10.1063/1.1487570} {\bibfield
  {journal} {\bibinfo  {journal} {Chaos}\ }\textbf {\bibinfo {volume} {12}},\
  \bibinfo {pages} {549--558} (\bibinfo {year} {2002})}\BibitemShut {NoStop}%
\bibitem [{\citenamefont {Packard}\ \emph {et~al.}(1980)\citenamefont
  {Packard}, \citenamefont {Crutchfield}, \citenamefont {Farmer},\ and\
  \citenamefont {Shaw}}]{Pac}%
  \BibitemOpen
  \bibfield  {author} {\bibinfo {author} {\bibfnamefont {N.}~\bibnamefont
  {Packard}}, \bibinfo {author} {\bibfnamefont {J.}~\bibnamefont
  {Crutchfield}}, \bibinfo {author} {\bibfnamefont {J.}~\bibnamefont {Farmer}},
  \ and\ \bibinfo {author} {\bibfnamefont {R.}~\bibnamefont {Shaw}},\
  }\bibfield  {title} {\enquote {\bibinfo {title} {Geometry from a time
  series},}\ }\href@noop {} {\bibfield  {journal} {\bibinfo  {journal}
  {Physical Review Letters}\ }\textbf {\bibinfo {volume} {45}},\ \bibinfo
  {pages} {712–716} (\bibinfo {year} {1980})}\BibitemShut {NoStop}%
\bibitem [{\citenamefont {Eckmann}, \citenamefont {{Oliffson Kamphorst}},\ and\
  \citenamefont {Ruelle}(1987)}]{Eckmann}%
  \BibitemOpen
  \bibfield  {author} {\bibinfo {author} {\bibfnamefont {J.-P.}\ \bibnamefont
  {Eckmann}}, \bibinfo {author} {\bibfnamefont {S.}~\bibnamefont {{Oliffson
  Kamphorst}}}, \ and\ \bibinfo {author} {\bibfnamefont {D.}~\bibnamefont
  {Ruelle}},\ }\bibfield  {title} {\enquote {\bibinfo {title} {{Recurrence
  Plots} of {Dynamical Systems}},}\ }\href {\doibase 10.1209/0295-5075/4/9/004}
  {\bibfield  {journal} {\bibinfo  {journal} {Europhysics Letters}\ }\textbf
  {\bibinfo {volume} {4}},\ \bibinfo {pages} {973--977} (\bibinfo {year}
  {1987})}\BibitemShut {NoStop}%
\bibitem [{\citenamefont {Mindlin}\ and\ \citenamefont {Gilmore}(1992)}]{Mind}%
  \BibitemOpen
  \bibfield  {author} {\bibinfo {author} {\bibfnamefont {G.~M.}\ \bibnamefont
  {Mindlin}}\ and\ \bibinfo {author} {\bibfnamefont {R.}~\bibnamefont
  {Gilmore}},\ }\bibfield  {title} {\enquote {\bibinfo {title} {Topological
  analysis and synthesis of chaotic time series},}\ }\href {\doibase
  10.1016/0167-2789(92)90111-Y} {\bibfield  {journal} {\bibinfo  {journal}
  {Physica D}\ }\textbf {\bibinfo {volume} {58}},\ \bibinfo {pages} {229--242}
  (\bibinfo {year} {1992})}\BibitemShut {NoStop}%
\bibitem [{\citenamefont {Zbilut}\ and\ \citenamefont {{Webber,
  Jr.}}(1992)}]{Zbi1}%
  \BibitemOpen
  \bibfield  {author} {\bibinfo {author} {\bibfnamefont {J.~P.}\ \bibnamefont
  {Zbilut}}\ and\ \bibinfo {author} {\bibfnamefont {C.~L.}\ \bibnamefont
  {{Webber, Jr.}}},\ }\bibfield  {title} {\enquote {\bibinfo {title}
  {Embeddings and delays as derived from quantification of recurrence plots},}\
  }\href {\doibase 10.1016/0375-9601(92)90426-M} {\bibfield  {journal}
  {\bibinfo  {journal} {Physics Letters A}\ }\textbf {\bibinfo {volume}
  {171}},\ \bibinfo {pages} {199--203} (\bibinfo {year} {1992})}\BibitemShut
  {NoStop}%
\bibitem [{\citenamefont {Zbilut}, \citenamefont {{Zald\'ivar-Comenges}},\ and\
  \citenamefont {Strozzi}(2002)}]{Zbi2}%
  \BibitemOpen
  \bibfield  {author} {\bibinfo {author} {\bibfnamefont {J.~P.}\ \bibnamefont
  {Zbilut}}, \bibinfo {author} {\bibfnamefont {J.-M.}\ \bibnamefont
  {{Zald\'ivar-Comenges}}}, \ and\ \bibinfo {author} {\bibfnamefont
  {F.}~\bibnamefont {Strozzi}},\ }\bibfield  {title} {\enquote {\bibinfo
  {title} {{Recurrence quantification based Liapunov exponents for monitoring
  divergence in experimental data}},}\ }\href {\doibase
  10.1016/S0375-9601(02)00436-X} {\bibfield  {journal} {\bibinfo  {journal}
  {Physics Letters A}\ }\textbf {\bibinfo {volume} {297}},\ \bibinfo {pages}
  {173--181} (\bibinfo {year} {2002})}\BibitemShut {NoStop}%
\bibitem [{\citenamefont {Thiel}\ \emph {et~al.}(2002)\citenamefont {Thiel},
  \citenamefont {Romano}, \citenamefont {Kurths}, \citenamefont {Meucci},
  \citenamefont {Allaria},\ and\ \citenamefont {Arecchi}}]{Thi1}%
  \BibitemOpen
  \bibfield  {author} {\bibinfo {author} {\bibfnamefont {M.}~\bibnamefont
  {Thiel}}, \bibinfo {author} {\bibfnamefont {M.~C.}\ \bibnamefont {Romano}},
  \bibinfo {author} {\bibfnamefont {J.}~\bibnamefont {Kurths}}, \bibinfo
  {author} {\bibfnamefont {R.}~\bibnamefont {Meucci}}, \bibinfo {author}
  {\bibfnamefont {E.}~\bibnamefont {Allaria}}, \ and\ \bibinfo {author}
  {\bibfnamefont {F.~T.}\ \bibnamefont {Arecchi}},\ }\bibfield  {title}
  {\enquote {\bibinfo {title} {{Influence of observational noise on the
  recurrence quantification analysis}},}\ }\href {\doibase
  10.1016/S0167-2789(02)00586-9} {\bibfield  {journal} {\bibinfo  {journal}
  {Physica D}\ }\textbf {\bibinfo {volume} {171}},\ \bibinfo {pages} {138--152}
  (\bibinfo {year} {2002})}\BibitemShut {NoStop}%
\bibitem [{\citenamefont {Schinkel}, \citenamefont {Dimigen},\ and\
  \citenamefont {Marwan}(2008)}]{schinkel2008}%
  \BibitemOpen
  \bibfield  {author} {\bibinfo {author} {\bibfnamefont {S.}~\bibnamefont
  {Schinkel}}, \bibinfo {author} {\bibfnamefont {O.}~\bibnamefont {Dimigen}}, \
  and\ \bibinfo {author} {\bibfnamefont {N.}~\bibnamefont {Marwan}},\
  }\bibfield  {title} {\enquote {\bibinfo {title} {Selection of recurrence
  threshold for signal detection},}\ }\href {\doibase
  10.1140/epjst/e2008-00833-5} {\bibfield  {journal} {\bibinfo  {journal}
  {European Physical Journal -- Special Topics}\ }\textbf {\bibinfo {volume}
  {164}},\ \bibinfo {pages} {45--53} (\bibinfo {year} {2008})}\BibitemShut
  {NoStop}%
\bibitem [{\citenamefont {Matassini}\ \emph {et~al.}(2002)\citenamefont
  {Matassini}, \citenamefont {Kantz}, \citenamefont {{Ho\l yst}},\ and\
  \citenamefont {Hegger}}]{Mat}%
  \BibitemOpen
  \bibfield  {author} {\bibinfo {author} {\bibfnamefont {L.}~\bibnamefont
  {Matassini}}, \bibinfo {author} {\bibfnamefont {H.}~\bibnamefont {Kantz}},
  \bibinfo {author} {\bibfnamefont {J.~A.}\ \bibnamefont {{Ho\l yst}}}, \ and\
  \bibinfo {author} {\bibfnamefont {R.}~\bibnamefont {Hegger}},\ }\bibfield
  {title} {\enquote {\bibinfo {title} {Optimizing of recurrence plots for noise
  reduction},}\ }\href {\doibase 10.1103/PhysRevE.65.021102} {\bibfield
  {journal} {\bibinfo  {journal} {Physical Review E}\ }\textbf {\bibinfo
  {volume} {65}},\ \bibinfo {pages} {021102} (\bibinfo {year}
  {2002})}\BibitemShut {NoStop}%
\bibitem [{\citenamefont {Marwan}\ \emph {et~al.}(2009)\citenamefont {Marwan},
  \citenamefont {Donges}, \citenamefont {Zou}, \citenamefont {Donner},\ and\
  \citenamefont {Kurths}}]{Marwan3}%
  \BibitemOpen
  \bibfield  {author} {\bibinfo {author} {\bibfnamefont {N.}~\bibnamefont
  {Marwan}}, \bibinfo {author} {\bibfnamefont {J.~F.}\ \bibnamefont {Donges}},
  \bibinfo {author} {\bibfnamefont {Y.}~\bibnamefont {Zou}}, \bibinfo {author}
  {\bibfnamefont {R.~V.}\ \bibnamefont {Donner}}, \ and\ \bibinfo {author}
  {\bibfnamefont {J.}~\bibnamefont {Kurths}},\ }\bibfield  {title} {\enquote
  {\bibinfo {title} {Complex network approach for recurrence analysis of time
  series},}\ }\href {\doibase 10.1016/j.physleta.2009.09.042} {\bibfield
  {journal} {\bibinfo  {journal} {Physics Letters A}\ }\textbf {\bibinfo
  {volume} {373}},\ \bibinfo {pages} {4246--4254} (\bibinfo {year}
  {2009})}\BibitemShut {NoStop}%
\bibitem [{\citenamefont {Donner}\ \emph {et~al.}(2011)\citenamefont {Donner},
  \citenamefont {Heitzig}, \citenamefont {Donges}, \citenamefont {Zou},
  \citenamefont {Marwan},\ and\ \citenamefont {Kurths}}]{Don}%
  \BibitemOpen
  \bibfield  {author} {\bibinfo {author} {\bibfnamefont {R.~V.}\ \bibnamefont
  {Donner}}, \bibinfo {author} {\bibfnamefont {J.}~\bibnamefont {Heitzig}},
  \bibinfo {author} {\bibfnamefont {J.~F.}\ \bibnamefont {Donges}}, \bibinfo
  {author} {\bibfnamefont {Y.}~\bibnamefont {Zou}}, \bibinfo {author}
  {\bibfnamefont {N.}~\bibnamefont {Marwan}}, \ and\ \bibinfo {author}
  {\bibfnamefont {J.}~\bibnamefont {Kurths}},\ }\bibfield  {title} {\enquote
  {\bibinfo {title} {{The Geometry of Chaotic Dynamics -- A Complex Network
  Perspective}},}\ }\href {\doibase 10.1140/epjb/e2011-10899-1} {\bibfield
  {journal} {\bibinfo  {journal} {European Physical Journal B}\ }\textbf
  {\bibinfo {volume} {84}},\ \bibinfo {pages} {653--672} (\bibinfo {year}
  {2011})}\BibitemShut {NoStop}%
\bibitem [{\citenamefont {Donges}\ \emph {et~al.}(2012)\citenamefont {Donges},
  \citenamefont {Heitzig}, \citenamefont {Donner},\ and\ \citenamefont
  {Kurths}}]{Donges2012}%
  \BibitemOpen
  \bibfield  {author} {\bibinfo {author} {\bibfnamefont {J.~F.}\ \bibnamefont
  {Donges}}, \bibinfo {author} {\bibfnamefont {J.}~\bibnamefont {Heitzig}},
  \bibinfo {author} {\bibfnamefont {R.~V.}\ \bibnamefont {Donner}}, \ and\
  \bibinfo {author} {\bibfnamefont {J.}~\bibnamefont {Kurths}},\ }\bibfield
  {title} {\enquote {\bibinfo {title} {Analytical framework for recurrence
  network analysis of time series},}\ }\href {\doibase
  10.1103/PhysRevE.85.046105} {\bibfield  {journal} {\bibinfo  {journal}
  {Physical Review E}\ }\textbf {\bibinfo {volume} {85}},\ \bibinfo {pages}
  {046105} (\bibinfo {year} {2012})}\BibitemShut {NoStop}%
\bibitem [{\citenamefont {Jacob}\ \emph {et~al.}(2016)\citenamefont {Jacob},
  \citenamefont {Harikrishnan}, \citenamefont {Misra},\ and\ \citenamefont
  {Ambika}}]{Jacob2016PRE}%
  \BibitemOpen
  \bibfield  {author} {\bibinfo {author} {\bibfnamefont {R.}~\bibnamefont
  {Jacob}}, \bibinfo {author} {\bibfnamefont {K.~P.}\ \bibnamefont
  {Harikrishnan}}, \bibinfo {author} {\bibfnamefont {R.}~\bibnamefont {Misra}},
  \ and\ \bibinfo {author} {\bibfnamefont {G.}~\bibnamefont {Ambika}},\
  }\bibfield  {title} {\enquote {\bibinfo {title} {Uniform framework for the
  recurrence-network analysis of chaotic time series},}\ }\href {\doibase
  10.1103/PhysRevE.93.012202} {\bibfield  {journal} {\bibinfo  {journal}
  {Physical Review E}\ }\textbf {\bibinfo {volume} {93}},\ \bibinfo {pages}
  {012202} (\bibinfo {year} {2016})}\BibitemShut {NoStop}%
\bibitem [{\citenamefont {Eroglu}\ \emph {et~al.}(2014)\citenamefont {Eroglu},
  \citenamefont {Marwan}, \citenamefont {Prasad},\ and\ \citenamefont
  {Kurths}}]{Ero}%
  \BibitemOpen
  \bibfield  {author} {\bibinfo {author} {\bibfnamefont {D.}~\bibnamefont
  {Eroglu}}, \bibinfo {author} {\bibfnamefont {N.}~\bibnamefont {Marwan}},
  \bibinfo {author} {\bibfnamefont {S.}~\bibnamefont {Prasad}}, \ and\ \bibinfo
  {author} {\bibfnamefont {J.}~\bibnamefont {Kurths}},\ }\bibfield  {title}
  {\enquote {\bibinfo {title} {Finding recurrence networks' threshold
  adaptively for a specific time series},}\ }\href {\doibase
  10.5194/npg-21-1085-2014} {\bibfield  {journal} {\bibinfo  {journal}
  {Nonlinear Processes in Geophysics}\ }\textbf {\bibinfo {volume} {21}},\
  \bibinfo {pages} {1085--1092} (\bibinfo {year} {2014})}\BibitemShut {NoStop}%
\bibitem [{\citenamefont {Wiedermann}\ \emph {et~al.}(2017)\citenamefont
  {Wiedermann}, \citenamefont {Donges}, \citenamefont {Kurths},\ and\
  \citenamefont {Donner}}]{Wie}%
  \BibitemOpen
  \bibfield  {author} {\bibinfo {author} {\bibfnamefont {M.}~\bibnamefont
  {Wiedermann}}, \bibinfo {author} {\bibfnamefont {J.~F.}\ \bibnamefont
  {Donges}}, \bibinfo {author} {\bibfnamefont {J.}~\bibnamefont {Kurths}}, \
  and\ \bibinfo {author} {\bibfnamefont {R.~V.}\ \bibnamefont {Donner}},\
  }\bibfield  {title} {\enquote {\bibinfo {title} {Mapping and discrimination
  of networks in the complexity-entropy plane},}\ }\href {\doibase
  10.1103/PhysRevE.96.042304} {\bibfield  {journal} {\bibinfo  {journal}
  {Physical Review E}\ }\textbf {\bibinfo {volume} {96}},\ \bibinfo {pages}
  {042304} (\bibinfo {year} {2017})}\BibitemShut {NoStop}%
\bibitem [{\citenamefont {Koebbe}\ and\ \citenamefont
  {Mayer-Kress}(1992)}]{Koe}%
  \BibitemOpen
  \bibfield  {author} {\bibinfo {author} {\bibfnamefont {M.}~\bibnamefont
  {Koebbe}}\ and\ \bibinfo {author} {\bibfnamefont {G.}~\bibnamefont
  {Mayer-Kress}},\ }\bibfield  {title} {\enquote {\bibinfo {title} {{Use of
  Recurrence Plots in the Analysis of Time-Series Data}},}\ }in\ \href@noop {}
  {\emph {\bibinfo {booktitle} {Proceedings of SFI Studies in the Science of
  Complexity}}},\ Vol.\ \bibinfo {volume} {XXI},\ \bibinfo {editor} {edited by\
  \bibinfo {editor} {\bibfnamefont {M.}~\bibnamefont {Casdagli}}\ and\ \bibinfo
  {editor} {\bibfnamefont {S.}~\bibnamefont {Eubank}}}\ (\bibinfo  {publisher}
  {Addison-Wesley},\ \bibinfo {address} {Redwood City},\ \bibinfo {year}
  {1992})\ pp.\ \bibinfo {pages} {361--378}\BibitemShut {NoStop}%
\bibitem [{\citenamefont {Zimek}, \citenamefont {Schubert},\ and\ \citenamefont
  {Kriegel}(2012)}]{Zimek}%
  \BibitemOpen
  \bibfield  {author} {\bibinfo {author} {\bibfnamefont {A.}~\bibnamefont
  {Zimek}}, \bibinfo {author} {\bibfnamefont {E.}~\bibnamefont {Schubert}}, \
  and\ \bibinfo {author} {\bibfnamefont {H.~P.}\ \bibnamefont {Kriegel}},\
  }\bibfield  {title} {\enquote {\bibinfo {title} {{A survey on unsupervised
  outlier detection in high-dimensional numerical data}},}\ }\href {\doibase
  10.1002/sam.11161} {\bibfield  {journal} {\bibinfo  {journal} {Statistical
  Analysis and Data Mining}\ }\textbf {\bibinfo {volume} {{5}}},\ \bibinfo
  {pages} {{363--387}} (\bibinfo {year} {{2012}})}\BibitemShut {NoStop}%
\bibitem [{\citenamefont {Marwan}\ \emph {et~al.}(2007)\citenamefont {Marwan},
  \citenamefont {Romano}, \citenamefont {Thiel},\ and\ \citenamefont
  {Kurths}}]{Marwan1}%
  \BibitemOpen
  \bibfield  {author} {\bibinfo {author} {\bibfnamefont {N.}~\bibnamefont
  {Marwan}}, \bibinfo {author} {\bibfnamefont {M.~C.}\ \bibnamefont {Romano}},
  \bibinfo {author} {\bibfnamefont {M.}~\bibnamefont {Thiel}}, \ and\ \bibinfo
  {author} {\bibfnamefont {J.}~\bibnamefont {Kurths}},\ }\bibfield  {title}
  {\enquote {\bibinfo {title} {{Recurrence Plots for the Analysis of Complex
  Systems}},}\ }\href {\doibase 10.1016/j.physrep.2006.11.001} {\bibfield
  {journal} {\bibinfo  {journal} {Physics Reports}\ }\textbf {\bibinfo {volume}
  {438}},\ \bibinfo {pages} {237--329} (\bibinfo {year} {2007})}\BibitemShut
  {NoStop}%
\bibitem [{\citenamefont {Sauer}, \citenamefont {J.A.Yorke},\ and\
  \citenamefont {Casdagli}(1991)}]{Sau}%
  \BibitemOpen
  \bibfield  {author} {\bibinfo {author} {\bibfnamefont {T.}~\bibnamefont
  {Sauer}}, \bibinfo {author} {\bibnamefont {J.A.Yorke}}, \ and\ \bibinfo
  {author} {\bibfnamefont {M.}~\bibnamefont {Casdagli}},\ }\bibfield  {title}
  {\enquote {\bibinfo {title} {Embedology},}\ }\href@noop {} {\bibfield
  {journal} {\bibinfo  {journal} {Journal of Statistical Physics}\ }\textbf
  {\bibinfo {volume} {65}},\ \bibinfo {pages} {579–616} (\bibinfo {year}
  {1991})}\BibitemShut {NoStop}%
\bibitem [{\citenamefont {Hegger}\ \emph {et~al.}(2000)\citenamefont {Hegger},
  \citenamefont {Kantz}, \citenamefont {Matassini},\ and\ \citenamefont
  {Schreiber}}]{Heg}%
  \BibitemOpen
  \bibfield  {author} {\bibinfo {author} {\bibfnamefont {R.}~\bibnamefont
  {Hegger}}, \bibinfo {author} {\bibfnamefont {H.}~\bibnamefont {Kantz}},
  \bibinfo {author} {\bibfnamefont {L.}~\bibnamefont {Matassini}}, \ and\
  \bibinfo {author} {\bibfnamefont {T.}~\bibnamefont {Schreiber}},\ }\bibfield
  {title} {\enquote {\bibinfo {title} {{Coping with Nonstationarity by
  Overembedding}},}\ }\href {\doibase 10.1103/PhysRevLett.84.4092} {\bibfield
  {journal} {\bibinfo  {journal} {Physical Review Letters}\ }\textbf {\bibinfo
  {volume} {84}},\ \bibinfo {pages} {4092--4095} (\bibinfo {year}
  {2000})}\BibitemShut {NoStop}%
\bibitem [{\citenamefont {Fraser}\ and\ \citenamefont
  {Swinney}(1986)}]{Fraser}%
  \BibitemOpen
  \bibfield  {author} {\bibinfo {author} {\bibfnamefont {A.~M.}\ \bibnamefont
  {Fraser}}\ and\ \bibinfo {author} {\bibfnamefont {H.~L.}\ \bibnamefont
  {Swinney}},\ }\bibfield  {title} {\enquote {\bibinfo {title} {{Independent
  coordinates for strange attractors from mutual information}},}\ }\href
  {\doibase 10.1103/PhysRevA.33.1134} {\bibfield  {journal} {\bibinfo
  {journal} {{Physical Review A}}\ }\textbf {\bibinfo {volume} {{33}}},\
  \bibinfo {pages} {{1134--1140}} (\bibinfo {year} {{1986}})}\BibitemShut
  {NoStop}%
\bibitem [{\citenamefont {Leadbetter}, \citenamefont {Lindgren},\ and\
  \citenamefont {Rootzen}(1983)}]{Leadbetter}%
  \BibitemOpen
  \bibfield  {author} {\bibinfo {author} {\bibfnamefont {M.~R.}\ \bibnamefont
  {Leadbetter}}, \bibinfo {author} {\bibfnamefont {G.}~\bibnamefont
  {Lindgren}}, \ and\ \bibinfo {author} {\bibfnamefont {H.}~\bibnamefont
  {Rootzen}},\ }\href@noop {} {\emph {\bibinfo {title} {{Extremes and Related
  Properties of Random Sequences and Processes}}}}\ (\bibinfo  {publisher}
  {Springer, New York},\ \bibinfo {year} {1983})\BibitemShut {NoStop}%
\bibitem [{\citenamefont {{Abarbanel}}(1996)}]{Abarbanel}%
  \BibitemOpen
  \bibfield  {author} {\bibinfo {author} {\bibfnamefont {H.~D.}\ \bibnamefont
  {{Abarbanel}}},\ }\href {\doibase 10.1007/978-1-4612-0763-4} {\emph {\bibinfo
  {title} {{Analysis of Observed Chaotic Data}}}}\ (\bibinfo  {publisher}
  {Springer, New York},\ \bibinfo {year} {1996})\BibitemShut {NoStop}%
\bibitem [{\citenamefont {Thiel}, \citenamefont {Romano},\ and\ \citenamefont
  {Kurths}(2006)}]{Thiel2006}%
  \BibitemOpen
  \bibfield  {author} {\bibinfo {author} {\bibfnamefont {M.}~\bibnamefont
  {Thiel}}, \bibinfo {author} {\bibfnamefont {M.~C.}\ \bibnamefont {Romano}}, \
  and\ \bibinfo {author} {\bibfnamefont {J.}~\bibnamefont {Kurths}},\
  }\bibfield  {title} {\enquote {\bibinfo {title} {Spurious structures in
  recurrence plots induced by embedding},}\ }\href {\doibase
  10.1007/s11071-006-2010-9} {\bibfield  {journal} {\bibinfo  {journal}
  {Nonlinear Dynamics}\ }\textbf {\bibinfo {volume} {44}},\ \bibinfo {pages}
  {299--305} (\bibinfo {year} {2006})}\BibitemShut {NoStop}%
\bibitem [{\citenamefont {{Lorenz}}(1963)}]{Lorenz}%
  \BibitemOpen
  \bibfield  {author} {\bibinfo {author} {\bibfnamefont {E.~N.}\ \bibnamefont
  {{Lorenz}}},\ }\bibfield  {title} {\enquote {\bibinfo {title} {Deterministic
  nonperiodic flow},}\ }\href {\doibase
  10.1175/1520-0469(1963)020<0130:DNF>2.0.CO;2} {\bibfield  {journal} {\bibinfo
   {journal} {Journal of the Atmospheric Sciences}\ }\textbf {\bibinfo {volume}
  {20}},\ \bibinfo {pages} {130--141} (\bibinfo {year} {1963})}\BibitemShut
  {NoStop}%
\bibitem [{\citenamefont {Ngamga}\ \emph {et~al.}(2016)\citenamefont {Ngamga},
  \citenamefont {Bialonski}, \citenamefont {Marwan}, \citenamefont {Kurths},
  \citenamefont {Geier},\ and\ \citenamefont {Lehnertz}}]{ngamga2016}%
  \BibitemOpen
  \bibfield  {author} {\bibinfo {author} {\bibfnamefont {E.~J.}\ \bibnamefont
  {Ngamga}}, \bibinfo {author} {\bibfnamefont {S.}~\bibnamefont {Bialonski}},
  \bibinfo {author} {\bibfnamefont {N.}~\bibnamefont {Marwan}}, \bibinfo
  {author} {\bibfnamefont {J.}~\bibnamefont {Kurths}}, \bibinfo {author}
  {\bibfnamefont {C.}~\bibnamefont {Geier}}, \ and\ \bibinfo {author}
  {\bibfnamefont {K.}~\bibnamefont {Lehnertz}},\ }\bibfield  {title} {\enquote
  {\bibinfo {title} {{Evaluation of selected recurrence measures in
  discriminating pre-ictal and inter-ictal periods from epileptic EEG data}},}\
  }\href {\doibase 10.1016/j.physleta.2016.02.024} {\bibfield  {journal}
  {\bibinfo  {journal} {Physics Letters A}\ }\textbf {\bibinfo {volume}
  {380}},\ \bibinfo {pages} {1419--1425} (\bibinfo {year} {2016})}\BibitemShut
  {NoStop}%
\bibitem [{\citenamefont {Little}\ \emph {et~al.}(2007)\citenamefont {Little},
  \citenamefont {{McSharry}}, \citenamefont {Roberts}, \citenamefont
  {Costello},\ and\ \citenamefont {Moroz}}]{little2007}%
  \BibitemOpen
  \bibfield  {author} {\bibinfo {author} {\bibfnamefont {M.~A.}\ \bibnamefont
  {Little}}, \bibinfo {author} {\bibfnamefont {P.~E.}\ \bibnamefont
  {{McSharry}}}, \bibinfo {author} {\bibfnamefont {S.~J.}\ \bibnamefont
  {Roberts}}, \bibinfo {author} {\bibfnamefont {D.~A.~E.}\ \bibnamefont
  {Costello}}, \ and\ \bibinfo {author} {\bibfnamefont {I.~M.}\ \bibnamefont
  {Moroz}},\ }\bibfield  {title} {\enquote {\bibinfo {title} {{Exploiting
  Nonlinear Recurrence and Fractal Scaling Properties for Voice Disorder
  Detection}},}\ }\href {\doibase 10.1186/1475-925X-6-23} {\bibfield  {journal}
  {\bibinfo  {journal} {BioMedical Engineering OnLine}\ }\textbf {\bibinfo
  {volume} {6}},\ \bibinfo {pages} {1--19} (\bibinfo {year}
  {2007})}\BibitemShut {NoStop}%
\bibitem [{\citenamefont {Baptista}\ \emph {et~al.}(2010)\citenamefont
  {Baptista}, \citenamefont {Ngamga}, \citenamefont {Pinto}, \citenamefont
  {Brito},\ and\ \citenamefont {Kurths}}]{baptista2010}%
  \BibitemOpen
  \bibfield  {author} {\bibinfo {author} {\bibfnamefont {M.~S.}\ \bibnamefont
  {Baptista}}, \bibinfo {author} {\bibfnamefont {E.~J.}\ \bibnamefont
  {Ngamga}}, \bibinfo {author} {\bibfnamefont {P.~R.~F.}\ \bibnamefont
  {Pinto}}, \bibinfo {author} {\bibfnamefont {M.}~\bibnamefont {Brito}}, \ and\
  \bibinfo {author} {\bibfnamefont {J.}~\bibnamefont {Kurths}},\ }\bibfield
  {title} {\enquote {\bibinfo {title} {{Kolmogorov-Sinai entropy from
  recurrence times}},}\ }\href {\doibase 10.1016/j.physleta.2009.12.057}
  {\bibfield  {journal} {\bibinfo  {journal} {Physics Letters A}\ }\textbf
  {\bibinfo {volume} {374}},\ \bibinfo {pages} {1135--1140} (\bibinfo {year}
  {2010})}\BibitemShut {NoStop}%
\bibitem [{\citenamefont {{Lekscha}}\ and\ \citenamefont
  {{Donner}}(2018)}]{Lekscha}%
  \BibitemOpen
  \bibfield  {author} {\bibinfo {author} {\bibfnamefont {J.}~\bibnamefont
  {{Lekscha}}}\ and\ \bibinfo {author} {\bibfnamefont {R.~V.}\ \bibnamefont
  {{Donner}}},\ }\bibfield  {title} {\enquote {\bibinfo {title} {{Phase space
  reconstruction for non-uniformly sampled noisy time series}},}\ }\href@noop
  {} {\bibfield  {journal} {\bibinfo  {journal} {ArXiv e-prints}\ } (\bibinfo
  {year} {2018})},\ \Eprint {http://arxiv.org/abs/1801.09517} {arXiv:1801.09517
  [physics.data-an]} \BibitemShut {NoStop}%
\bibitem [{\citenamefont {Freedman}\ and\ \citenamefont
  {Diaconis}(1981)}]{Freedman}%
  \BibitemOpen
  \bibfield  {author} {\bibinfo {author} {\bibfnamefont {D.}~\bibnamefont
  {Freedman}}\ and\ \bibinfo {author} {\bibfnamefont {P.}~\bibnamefont
  {Diaconis}},\ }\bibfield  {title} {\enquote {\bibinfo {title} {{On the
  Histogram as a Density Estimator: $L_2$ Theory}},}\ }\href {\doibase
  10.1007/BF01025868} {\bibfield  {journal} {\bibinfo  {journal} {Zeitschrift
  f\"ur Wahrscheinlichkeitstheorie und Verwandte Gebiete}\ }\textbf {\bibinfo
  {volume} {57}},\ \bibinfo {pages} {453--476} (\bibinfo {year}
  {1981})}\BibitemShut {NoStop}%
\end{thebibliography}%


\appendix
\renewcommand{\thefigure }{A\arabic{figure}}
\setcounter{figure}{0}

\section{Influence of embedding dimension on the variations in the maximum and 
mean pairwise distances}\label{S_influence}

\begin{figure}
 \centering
 \includegraphics[scale=0.31]{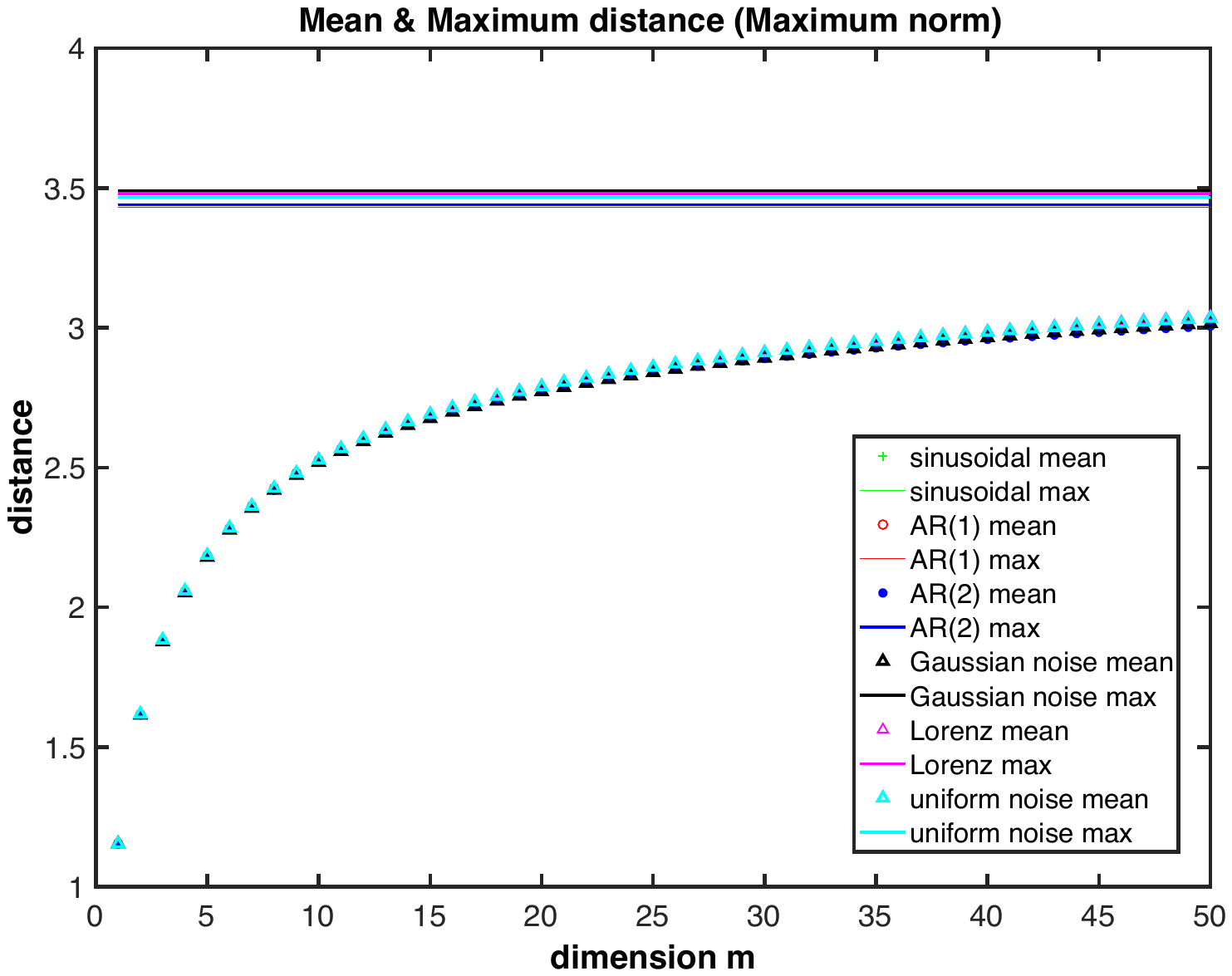}
 \caption{Mean $d^{(\infty)}_{mean}$ and maximum $d^{(\infty)}_{max}$ $L_\infty$ distance between all pairs of state vectors as a function 
of the embedding dimension $m$ for different types of time series: polychromatic harmonic 
oscillation with periods 3, 50 and 500; auto-regressive processes of first and second order 
with parameters $\varphi_1 = 0.5, \varphi_2 = 0.3$; random numbers of standard Gaussian (zero 
mean and unit variance) and uniform (unit variance) distributions, and $y$ component of the 
Lorenz-63 system (Eq.~\eqref{Lorenz1}, see Section \ref{numerical}) with control parameters $\sigma=10$, $\beta=8/3$ and $r$ linearly increasing from 180 (chaotic regime) to 210 (periodic regime) as a function of the embedding dimension $m$.}
\label{S_figure1}
\end{figure}

\begin{figure}
 \centering
 \includegraphics[scale=0.31]{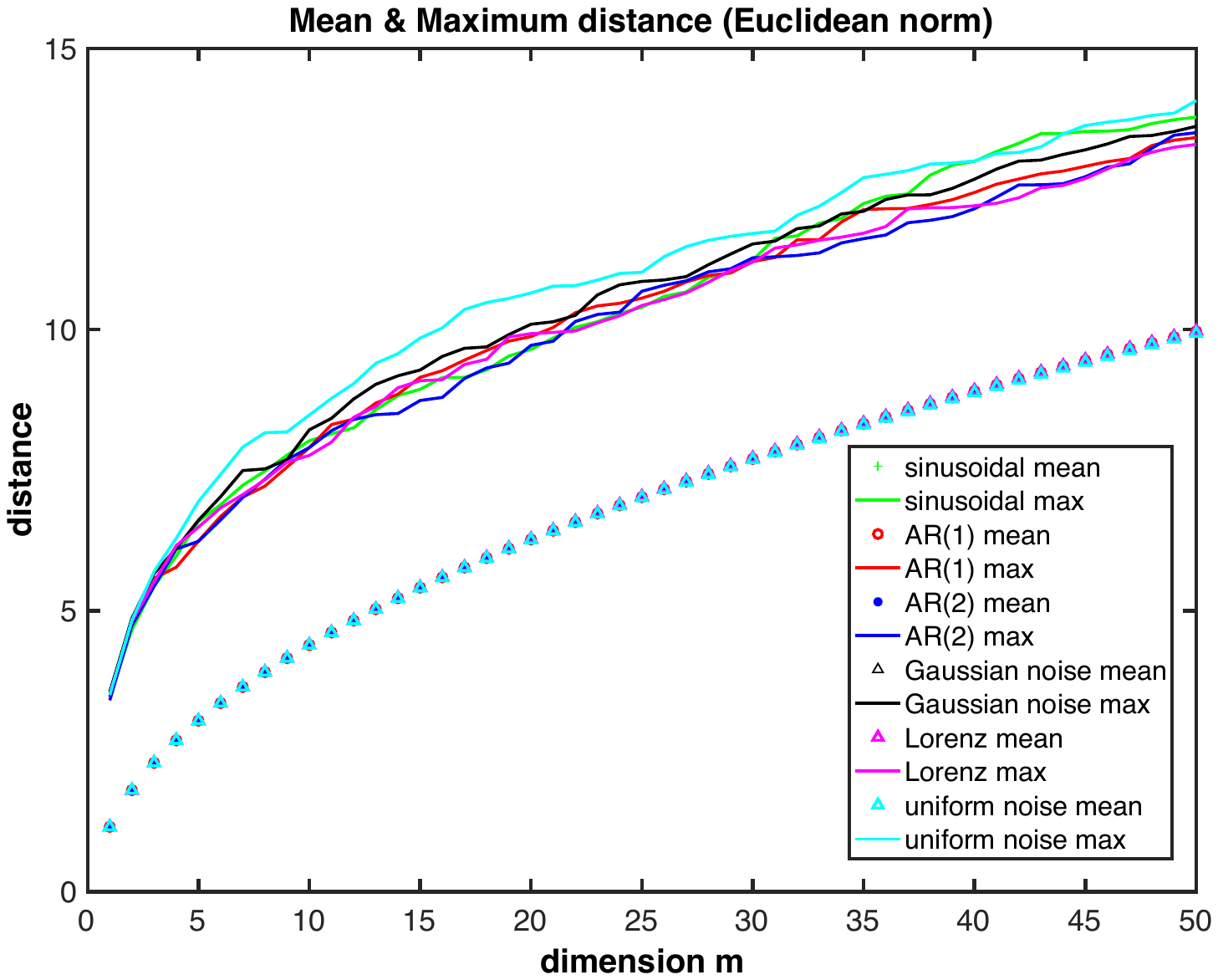}
 \caption{Same as in Fig.~\ref{S_figure1} for \textit{$L_2$} distances.}
 \label{S_figure2}
\end{figure}

As discussed in Section~\ref{influence}, we show some numerical results illustrating the general behavior of mean and maximum $L_\infty$ and $L_2$ distances for different types of systems in Figs.~\ref{S_figure1} and \ref{S_figure2}, respectively. For a theoretical explanation of the observed changes with increasing embedding dimension, see Section~\ref{influence}.

\section{Empirical shape parameters of the distance distributions for different systems}\label{S_variations}

In order to further characterize the shape of the empirically observed pairwise distance distributions shown in Fig.~\ref{dist_examples}
in more detail, we consider two standard characteristics from descriptive statistics. On the one hand, the skewness
\begin{equation}
\hat{s} = \frac{\frac{1}{N_d}\sum_{i=1}^{N_d}(d_i-\bar{d})^3}
{\left( \sqrt{\frac{1}{N_d}\sum_{i=1}^{N_d}
 (d_i - \bar{d})^2} \right)^3} \label{skewness}
\end{equation}
of the distribution measures its asymmetry around the sample mean distance \textit{$\bar{d}$}.
On the other hand, we study the associated Shannon entropy
\begin{equation}
\label{shannon} \hat{h} = -\sum_{j=1}^{N_b} p_j
\frac{\log(p_j)}{\log(N_b)}
\end{equation}
providing an integral measure of the heterogeneity of the distribution of $d$. Here, $j$ enumerates the bins of a histogram of the values of $d$ with $N_b$ bins and relative frequencies $p_j$, and $N_d$ is 
the number of pairwise distances in the sample (i.e., the number of independent entries of the 
distance matrix $\mathbf{d}$, $N_d=N_{eff}(N_{eff}-1)/2$). 
The bin width has been selected by first computing the optimum value according to the Freedman-Diaconis rule\cite{Freedman} for each embedding dimension $m$ and then averaging over all corresponding values and taking the resulting mean to keep $N_b$ fixed for each considered setting. Specifically, for the time series drawn from the Gaussian and uniform distributions, $N_{b,L_2}=355$ and $N_{b,L_{\infty}}=286$, while for the Lorenz system,  $N_{b,L_2}=701$ and $N_{b,L_{\infty}}=771$.

According to the corresponding normalization, $\hat{h}$ 
assumes its maximum of one in case of a uniform distribution (since then, $p_j=1/N_b \quad \forall\, j=1,...,N_b$, i.e., for each (binned) distance within $[d_{min},d_{max}]$). In turn, the more heterogeneous (e.g., spiky or generally asymmetric) the distribution of distances gets, the lower $\hat{h}$.

\begin{figure*}[!h]
 \centering
 \includegraphics[width=0.9\textwidth]{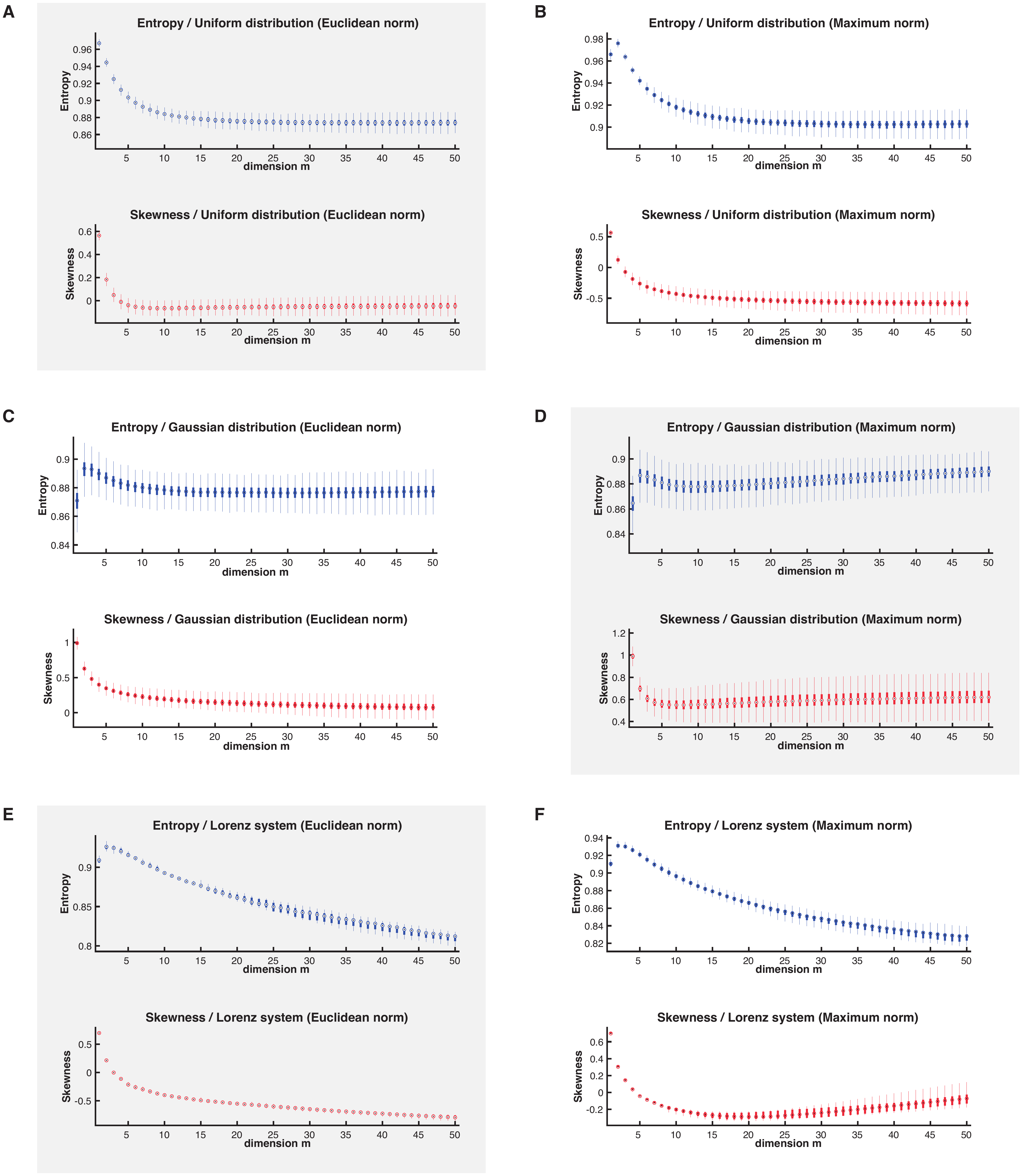}
\caption{Skewness (red) and Shannon entropy (blue) of the $L_2$ (A,C,E) and $L_\infty$ (B,D,F) distances of $N=1,500$ independent random numbers with uniform (A,B) and Gaussian (C,D) distribution and (E,F) the $y$ component of the Lorenz-63 system (Eq.~\eqref{Lorenz1}, $N=6,000$, see Section \ref{numerical}) with control parameters $\sigma=10$, $\beta=8/3$ and $r$ linearly increasing from 180 (chaotic regime) to 210 (periodic regime) as a function of the embedding dimension $m$. For the two noise series, box plots show the variability estimated from 1,000 independent realizations for each data set, using a random number generator. In case of the Lorenz-63 system the variability is estimated from 10 independent realizations of the non-stationary Lorenz-63 equations with randomly chosen initial conditions.}
\label{S_statistics}
\end{figure*}

Figure~\ref{S_statistics} shows the resulting behavior of both characteristics for the
$L_2$ (panels A,C,E) and $L_\infty$ (panels B,D,F) distances obtained from uniform and
Gaussian distributed noise as well as for the non-stationary Lorenz-63 system (Eq.~\eqref{Lorenz1},
see Section~\ref{numerical}) in dependence on the embedding dimension. The results
complement the qualitative description based on a visual inspection of Fig.~\ref{dist_examples} as given in Section~\ref{influence}. In case of the
$L_2$ norm and time series drawn from uniform and Gaussian distributions (Fig.~\ref{S_statistics} A,C) we observe the skewness converging towards zero (symmetric Gaussian
distribution) and the entropy reflecting this convergence towards a normal distribution
by a downward trend until the skewness approaches zero as $m$ further increases.
Although the theoretically predicted Gaussian shape for high $m$ is visually apparent
in case of the time series from the Lorenz-63 system (see Fig.~\ref{dist_examples} E), the skewness takes clearly non-zero negative values while the entropy constantly decreases with increasing $m$, indicating an asymmetric shape (Fig.~\ref{S_statistics} E). In case of the $L_\infty$ norm, the considered maximum embedding dimension appears not suited for observing convergence of both shape parameters.

\section{RP's and RQA for one realization of the non-stationary Lorenz system}\label{S_Lorenz_exp}

For further illustrating the RPs resulting from the time-dependent Lorenz-63 system discussed in Section~\ref{numerical}, we show here the results for just one example trajectory corresponding to a set of randomly chosen initial conditions $x(0)=0.9649$, $y(0)=0.1576$, $z(0)=0.9706$. As before, we embed the $y$ component time series and study the RP for each previously discussed threshold selection method. Then, we use a running window over each (global) RP with a window size of $w=400$ and mutual shift of $ws=40$ data points, i.e., 90\% overlap between consecutive windows. 

\begin{figure*}
 \centering
 \includegraphics[width=0.95\textwidth]{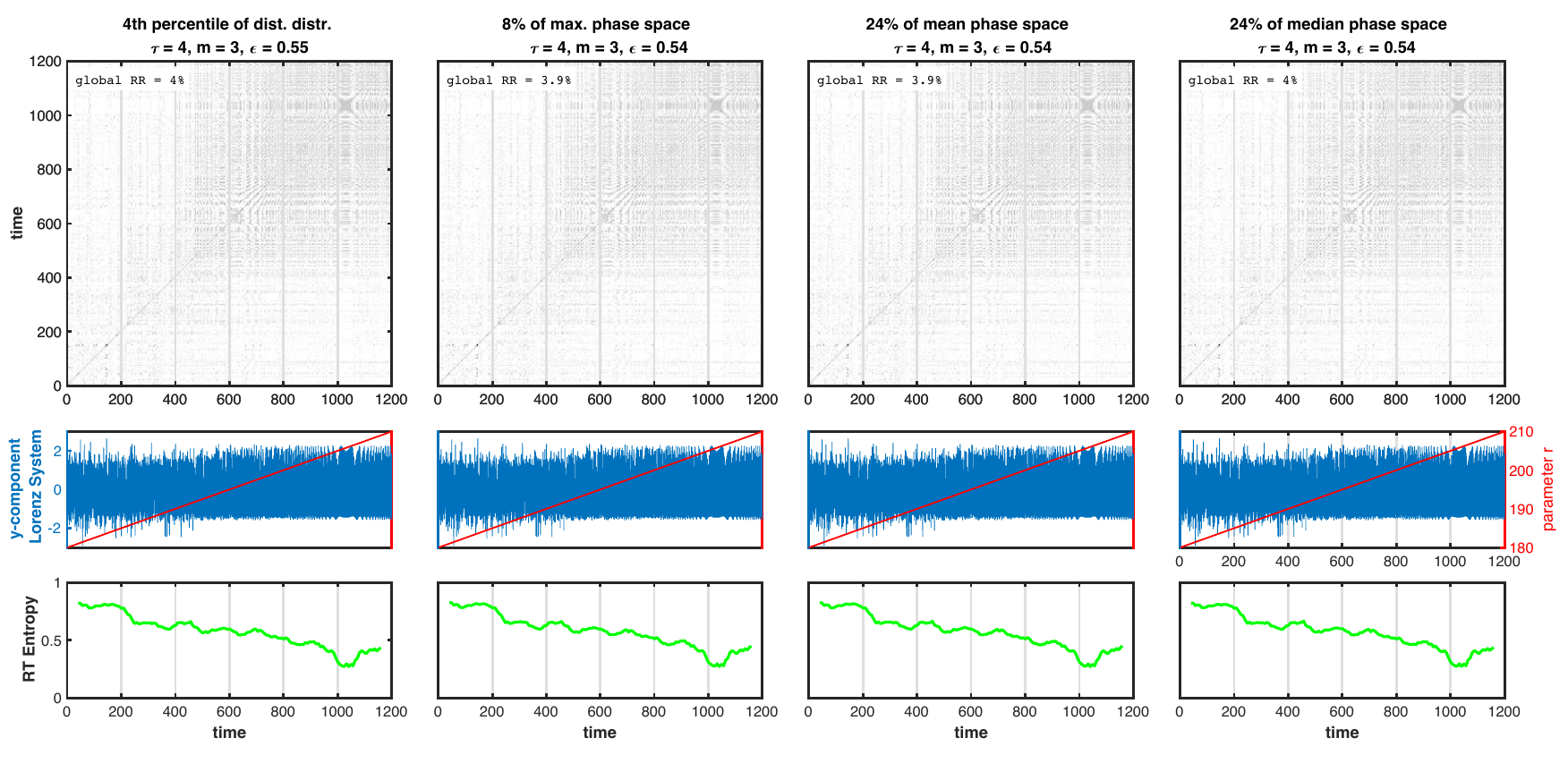}
  \caption{RPs, according time series (blue), time-dependence of the control parameter r (red) and recurrence characteristic $RTE$ (green) based on the $y$ component of the non-stationary Lorenz-63 system (see text for details), using the $L_2$ norm. Shown are the results for low-dimensional embedding ($m=3$) and for four different methods to select the recurrence threshold according to a certain percentile of the distance distribution and some percentage of the maximum, mean or median distances of state vectors on the reconstructed attractor (from left to right). The actual threshold values ($4$th percentile, $8\%$, $24\%$ and $24\%$, respectively) have again been chosen such that the global recurrence rate for each method in this embedding scenario is $\approx 4\%$.}
\label{S_results_Lorenz3}
\end{figure*} 

The RPs and the associated time-dependent recurrence characteristic $RTE$ (Eq.~\eqref{eq_rte}) for a ``normal'' three-dimensional embedding with time delay $\tau=4$, consistent with the first local minimum of the mutual information\cite{Fraser}, are shown in Fig.~\ref{S_results_Lorenz3}, using the Euclidean norm. We compare the results for four different threshold selection methods but similar effective threshold values (corresponding to a global recurrence rate of $RR\approx 4\%$), which are thus expected to give comparable results. The left panel corresponds to the recommended method of taking a certain percentile of the distance distribution, while the other three panels are based on thresholds selected according to some percentage of the maximum, mean and median distance of state vectors on the attractor in the reconstructed state space. Comparing the different panels, as expected there are hardly any marked differences in the RPs or the temporal changes of $RTE$. The transition from a chaotic regime into a periodic one is well reflected by a constantly decreasing $RTE$, which takes its minimum for the limit cycle behavior between $t_1\approx 1,000$ and $t_2\approx 1,080$. 

\begin{figure*}
 \centering
 \includegraphics[width=0.95\textwidth]{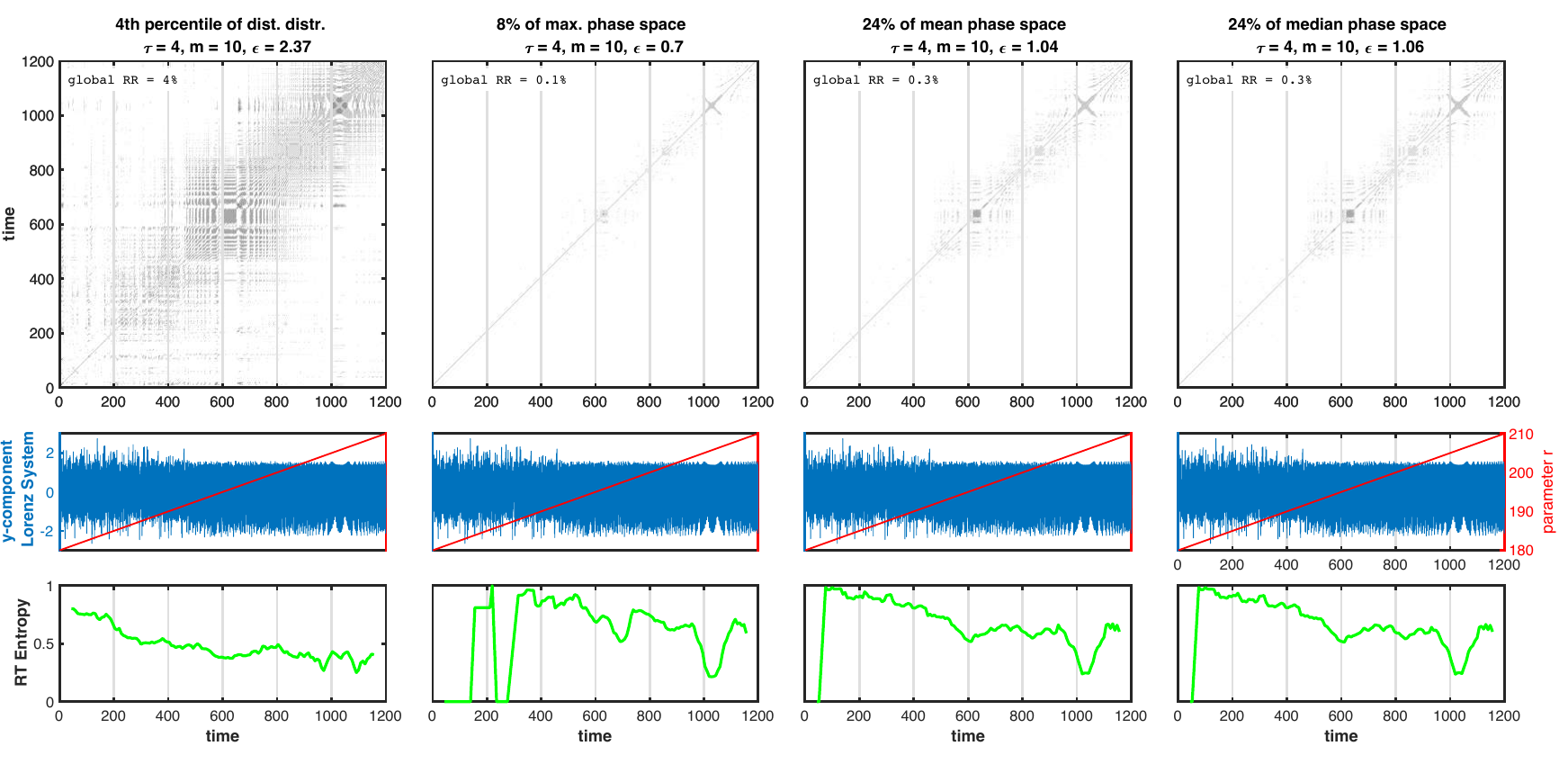}
\caption{Same as in Fig.~\ref{S_results_Lorenz3}, but for ten-dimensional embedding. In comparison to Fig.~\ref{S_results_Lorenz3}, three of the four methods lead to  a marked drop in the global recurrence rate and a resulting change in the $RTE$ values. Only for a recurrence threshold corresponding to the same percentile of the distance distribution, the results are qualitatively stable over the full considered time evolution.}
\label{S_results_Lorenz10}
\end{figure*} 

However, if choosing a higher-dimensional embedding (e.g., $m=10$) motivated by the 
non-stationarity of the system, the RP becomes almost completely white if the recurrence threshold is chosen based upon the same percentages of the maximum, mean or median state space distances as used before (Fig.~\ref{S_results_Lorenz10}). In this case the $RTE$ is still able to detect the transitory limit cycle regime, but one looses information about the chaotic regime before. In contrast, we retain the same density of recurrences and, hence, resolution of the RP as for $m=3$ when fixing the threshold according to the whole distance distribution (left panel in Fig.~\ref{S_results_Lorenz10}). Here, the overall behavior of $RTE$ from the lower-dimensional ($m=3$) case is qualitatively retained, although the periodic regime is less well expressed than in the former case.

\end{document}